\documentclass[%
 reprint,
 preprintnumbers,
 amsmath,amssymb,
 aps,
 floatfix,
]{revtex4-2}

\usepackage{graphicx}
\usepackage{grffile}
\usepackage{dcolumn}
\usepackage{bm}
\usepackage{xspace}
\usepackage{ifthen}
\usepackage{xifthen}
\usepackage{afterpage}
\usepackage{delimset}
\usepackage{stmaryrd}
\usepackage{siunitx}
\usepackage{color}
\usepackage{tikz}
\usetikzlibrary{arrows}
\usetikzlibrary{calc}

\newcommand{\fvec}[1]         {\boldsymbol{#1}}

\newcommand{\diff}            {\operatorname{d}\!}
\newcommand{\prob}       [2][]{%
  \ensuremath{%
    \mathbb{P}\!\ifthenelse{\isempty{#1}}%
      {\left[{#2}\right]}%
      {\left[{#2}\,\middle|\,{#1}\,\right]}%
  }\xspace%
}

\newcommand{\YL}              {Y_{\rm L}}
\newcommand{\Afrag}           {A_{\rm f}}
\newcommand{\Zfrag}           {Z_{\rm f}}
\newcommand{\Nfrag}           {N_{\rm f}}
\newcommand{\Xfrag}           {X_{\rm f}}
\newcommand{\MQS}        [1]  {\fvec{#1}}

\newcommand{\ket}        [1]  {\ensuremath{\left|{#1}\right\rangle}\xspace}
\newcommand{\braket}     [3][]{%
  \ensuremath{
    \ifthenelse{\isempty{#1}}{%
      \left\langle{#2}\middle|{#3}\right\rangle%
    }{%
      \delimtriple<\vert\vert>!{#2}{#1}{#3}%
    }
  }\xspace%
}

\newcommand{\shapeInside} [1] {\Omega}

\newcommand{\ladder} [4]{\def\tempid{#1}\def\tempac{#2}\ifx\tempid\empty {#3}_{#4}\ifx#2c^{\dagger}\fi \else {#3}^{(#1)\ifx#2c{\dagger}\fi}_{#4}
                          \fi}
\newcommand{\idPw}      {a}

\newcommand{\cPw} [2][] {\ladder{#1}{c}{\hat{\idPw}}{#2}}

\newcommand{\timerev}[1]{\bar{#1}}

\newcommand{\figref}       [1]{Fig.~\ref{#1}}
\newcommand{\secref}       [1]{Sec.~\ref{#1}}
\newcommand{\subsecref}    [1]{Sec.~\ref{#1}}

\tikzset{
    partial ellipse/.style args={#1:#2:#3}{
        insert path={+ (#1:#3) arc (#1:#2:#3)}
    }
}

\begin{document}

\preprint{LA-UR-20-26369}
\preprint{LLNL-JRNL-813628}

\title{Improvements to the macroscopic-microscopic approach of nuclear fission}

\author{Marc Verriere}%
 \email{verriere1@llnl.gov}
 \affiliation{%
 Theoretical Division, Los Alamos National Laboratory, Los Alamos, NM 87545, USA \\
 Nuclear and Chemical Sciences Division, Lawrence Livermore National Laboratory, Livermore, California 94551, USA
 }
\author{Matthew Ryan Mumpower}%
 \email{mumpower@lanl.gov}
\affiliation{%
Theoretical Division, Los Alamos National Laboratory, Los Alamos, NM 87545, USA
}%

\date{\today}

\begin{abstract}
The well established macroscopic-microscopic (mac-mic) description of nuclear fission enables the prediction of fission fragment yields for a broad range of fissioning systems. 
In this work, we present several key enhancements to this approach. 
We improve upon the microscopic sector of nuclear potential energy surfaces by magnifying the resolution of the Lipkin-Nogami equations and strengthening the Strutinsky procedure, thus reducing spurious effects from the continuum. 
We further present a novel deterministic method for calculating fission dynamics under the assumption of strongly damped nucleonic motion. 
Our technique directly determines the evolution of the scissioned shape distribution according to the number of random walk steps rather than the statistical accumulation of fission events. 
We show that our new technique is equivalent to the Metropolis random walk pioneered over the past decade by Randrup and colleagues.
It further improves upon it, as we remove the need for altering the nuclear landscape via a biased potential. 
With our final improvement, we calculate fission fragments mass and charge distributions using particle number projection, which affords the simultaneous calculation of both mass and charge yield distributions.
Fission fragments are thus calculated from the quantum mechanical $A$-body states of the potential energy surface rather than the collective mass asymmetry variable ($\alpha_{\rm g}$) of the Finite-Range Liquid-Drop Model (FRLDM) used in past work. 
We highlight the success of our enhancements by predicting the odd-even staggering and the charge polarization for the neutron-induced fission of ${}^{233}$U and ${}^{235}$U. 
\end{abstract}

\maketitle



\section{Introduction}
\label{sec:intro}

Fission data, especially those based on fragment yields, are a key ingredient in many applications.
For instance, the accurate description of fission yields is influential for nuclear engineering and waste management \cite{tsoulfanidis1991wm, nrp2018}, the production of radioactive isotopes via fragmentation \cite{pardo2007frag, carlson2017frag, sherrill2018frag, surman2018frag, abel2019frag}, reactor neutrinos \cite{mention2011rn, hayes2014rn}, and in the pursuit of synthesizing superheavy elements \cite{itkis2004she, oganessian2015she, godby202she, simenel2012she}.
Fission may also play an important role in the formation of the heavy elements in astrophysical processes \cite{martinez2007role, goriely2013new, eichler2015role, mumpower2018beta, vassh2019using, vassh2020frldm}.
The complexity of nuclear fission makes this reaction challenging to model theoretically from fundamental principles.

There are many approaches to the theoretical description of fission. 
Fully microscopic models describe the fission process by assuming the nuclear interaction between the nucleons only. 
The most frequently used implementation is the Energy Density Functional (EDF) theory~\cite{schunck2016microscopic}, where the density-dependence of the energy is derived from an effective interaction or directly parameterized~\cite{stone2007skyrme,robledo2018mean}.
Among EDF approaches, the time-dependent Hartree-Fock (in its broad sense, possibly including symmetry restorations or pairing correlations) consists in the explicit determination of the time-dependent evolution of the mean-field potential~\cite{engel1975time, bonche1976one, cusson1979time, levit1980barrier, simenel2014formation, goddard2015fission, bulgac2016induced, simenel2018heavy, tanimura2017microscopic, regnier2019microscopic}.
Developments are also being made to include two-body dissipation.
A first method, TD2RDM, is based on the time-dependent version of the reduced density matrix method and explicitly considers the two-body density matrix time-evolution~\cite{czuba2020combining}.
Another promising way is the inclusion of stochastic fluctuations in the dynamics~\cite{ayik1988fluctuations,randrup1990fluctuations,abe1996stochastic}.
A different strategy to describe fission involves using the time-dependent generator coordinate method (TDGCM) which is based on the assumption that only a few collective variables drive the overall fission dynamics. 
This hypothesis enables the reduction of the many degrees of freedom involved in the fissioning system description to only a couple of them (one to three in practical applications)~\cite{verriere2020timedependent, berger1991time, goutte2005microscopic, younes2009microscopic, schunck2016microscopic, regnier2016fission, regnier2019from, younes2019microscopic, zhao2019microscopic}. 
Efforts have been made to extend this approach and include dissipative effects in a microscopic framework~\cite{bernard2011microscopic, bernard2011taking, younes2019microscopic}.

Approaching the description of fission observables from macroscopic-microscopic theory provides yet another alternative. In this approach, the fissioning system is modeled to first approximation as a macroscopic system (e.g., a liquid-drop or a droplet of nuclear matter) to obtain the smooth part of its energy~\cite{moller2012frdm, moller2016nuclear}. Corrections are then applied that account for the missing microscopic behavior such as shell and pairing effects~\cite{strutinsky1967sc, strutinsky1968sc, brack1972sc}. The time-evolution of the fissioning system is then obtained statistically with the exact or approximate resolution of the Langevin equations~\cite{wada1992lang, nadtochy2007lang, sierk2017lang}. Approximate methods that assume strong damping are modeled via a random walk on the potential energy surface and are found to produce a good agreement with known data \cite{randrup2011brownian, randrup2013energy}. 

The benefit of such semi-classical approaches over the microscopic approaches applied to the description of fission is the inclusion of larger spaces for the collective shape degrees of freedom and the ability to model stochastic dynamics, while retaining reasonable calculation times. Thus, such models are applicable to a range of reactions across many heavy fissioning systems~\cite{mumpower2020primary}. To date, these approaches have examined several starting configurations for the dynamics, including from the ground state, inner saddle region, and after the last saddle \cite{sierk2017lang, mumpower2020primary}. Those studies which start after the last saddle explicitly ignore dynamics before this point, potentially neglecting important pathways. Studies that start at the ground state may introduce a phenomenological tilting of the potential energy surface in an attempt to force the calculation to go over the fission barrier sufficiently quickly. The downside to the tilting is that it may introduce an unphysical bias to the results. Finally, the fragments properties of past work, e.g., \cite{randrup2011brownian}, may be calculated with reference only to the collective asymmetry coordinate ($\alpha_{\rm g}$), thereby limiting the predictive power of the approach. 

In this work, we address several of the major shortcomings of past macroscopic-microscopic fission studies to describe the nascent fission fragment yields. 
In Sec.~\ref{sec:appr}, we outline our theoretical model and highlight the various improvements. 
In Sec.~\ref{sec:appli}, we present the results we have obtained with our approach. 

\section{Theoretical approach}
\label{sec:appr}

Our description of the fission process can be decomposed into three main steps. 
First, the shape-dependent potential energy surface (PES) of the corresponding fissioning system (typically target-plus-neutron) is determined using a semi-classical method based on the macroscopic Finite-Range Liquid-Drop Model (FRLDM) with microscopic corrections. 
The calculation of the potential energy surface, which has been improved by using an enhanced Strutinsky method that drastically reduces the continuum effects and allows the use of higher-quality particle basis, is presented in Sec.~\ref{subsec:appr:PES}. 
Second, we simulate the fission dynamics using a new algorithm presented in Sec.~\ref{subsec:appr:DPS}. 
This algorithm is used to obtain the probability to populate each scissioned configuration of the PES. 
The term ``scissioned configuration'' refers to the set of all the shapes where it is meaningful to calculate fission fragment properties.
Eventually, in what follows, such configurations are defined as the ones having a neck between the prefragments whose radius is below a threshold value $r_{\rm neck}$. 
Finally, in Sec.~\ref{subsec:appr:frag}, we present how we can deduce, for the first time in this type of approach, the fission fragment probability distribution in charge and mass, $Y(Z,A)$, based on a microscopic projection technique. 

\subsection{Potential energy landscape}
\label{subsec:appr:PES}

Several parameterizations of the sharp macroscopic density have been developed, see e.g., Ref.~\cite{hasse1988geometrical}, to specify the relevant degrees of freedom associated with large deformations encountered in fission. 
In this work, we restrict ourselves to the description of binary fission. Even though there is experimental evidence for ternary~\cite{ramayya1998cold, ramayya1998observation, vermote2008ternary, vermote2010ternary} and even quaternary~\cite{go2004ternary} fission, their contribution to the fragment probability distribution is smaller than binary fission by orders of magnitude. Thus, we use the so-called Matched-Quadratic-Surface (MQS) parameterization that was introduced in Refs.~\cite{nix1968further, nix1969further, nix1972calculation, bolsterli1972new} for the specification of our shape families. The MQS parameterization contains nine degrees of freedom. Six degrees of freedom remain by ensuring a smooth junction between the bodies and fixing the volume of the shape. Because one of the parameters corresponds to the center of mass, this may be set to the origin, reducing the number of parameters to five. We use the symbol $\MQS{\chi}$ to denote a specific MQS shape. The predefined grid we use to calculate the PES is taken from Ref.~\cite{moller2009heavy}. We note that the nodes associated with $\alpha_g=-0.02$ are redundant due to parity-reversal symmetry, and therefore are not explicitly calculated. 

For a nuclear system defined by $Z$ protons, and $N$ neutrons, the potential energy $E(\MQS{\chi})$ associated with a given set of MQS parameters can be written as
\begin{equation}\label{eq:appr:pes:Etot}
    E(\MQS{\chi}) =
        E_{\rm mac}(\MQS{\chi})
      + \Delta E_{\rm shell}(\MQS{\chi})
      + \Delta E_{\rm pair.}(\MQS{\chi}),
\end{equation}
where the first term, $E_{\rm mac}(\MQS{\chi})$, is the macroscopic energy obtained assuming that the fissioning system is a nuclear drop of charged liquid. It depends implicitly on $Z$ and $N$ as discussed in greater detail in Appendix~\ref{app:FRLDM}. The two remaining terms define the microscopic corrections.

In order to obtain the sharp contribution to the energy coming from shell effects, $\Delta E_{\rm shell}(\MQS{\chi})$, we determine a microscopic many-body state at the mean-field approximation. The effective averaged potential for the isospin $\tau$ is 
\begin{equation}\label{eq:appr:pes:micV}
    V^{(\tau)}(\MQS{\chi}) =
      V^{(\tau)}_{1}(\MQS{\chi})
    + V^{(\tau)}_{\rm C}(\MQS{\chi})
    + V^{(\tau)}_{\rm s.o.}(\MQS{\chi}),
\end{equation}
where each term is taken from Ref.~\cite{bolsterli1972new}. The first term, $V^{(\tau)}_{1}$, corresponds to the parameterized mean-field associated with the central part of the nuclear interaction and is obtained assuming a Yukawa interaction between the nucleons, 
\begin{equation}\label{eq:appr:pes:V1}
    V^{(\tau)}_{1}(\fvec{r}; \MQS{\chi}) = 
      -\frac{V_\tau}{4\pi a_{\rm pot}^3}
      \int_{\shapeInside{\chi}}\diff{\fvec{r}'}
        \frac{
          e^{-|\fvec{r}-\fvec{r}'|/a_{\rm pot}}
        }{
          |\fvec{r}-\fvec{r}'|/a_{\rm pot}
        },
\end{equation}
where $\shapeInside{\chi}$ is the volume delimited by the shape associated with the MQS parameters, $\MQS{\chi}$, scaled to have fixed volume of $\frac{4}{3}\pi A R_{\rm pot}^3$ ($R_{\rm pot}$ is defined by Eq.~(81) of Ref.~\cite{moeller1995nuclear}). The potential depths $V_{\tau}$ are given by
\begin{align}
    V_{\rm n} &= V_{\rm s} + V_{\rm a} \bar{\delta} \\
    V_{\rm p} &= V_{\rm s} - V_{\rm a} \bar{\delta},
\end{align}
where $V_{\rm s}$ and $V_{\rm a}$ are parameters of the model and $\bar{\delta}$ is given by Eq.~(85) of Ref.~\cite{moeller1995nuclear}.
The Coulomb term $V^{(\tau)}_{\rm C}(\fvec{r}; \MQS{\chi})$, is only acting on protons and, using the superposition principle, is given by
\begin{equation}
    V^{(\rm p)}_{\rm C}(\fvec{r}; \MQS{\chi}) =
      \int \diff{\fvec{r}'}\ 
        \rho^{(\MQS{\chi})}_{\rm prot.}(\fvec{r}')
        \frac{e^2}{|\fvec{r}-\fvec{r}'|},
\end{equation}
where $\rho_{\rm prot.}^{(\MQS{\chi})}(\fvec{r}_1)$ is the proton density. In our model, it is assumed sharp and homogeneous inside the shape, leading to
\begin{equation}\label{eq:appr:pes:C}
    V^{(\rm p)}_{\rm C}(\fvec{r}; \MQS{\chi}) =
      e \rho_{\rm c}
      \int_{\shapeInside{\chi}}\frac{\diff{\fvec{r}'}}{|\fvec{r}-\fvec{r}'|},
\end{equation}
where
\begin{equation}
    \rho_{\rm c} = \frac{Ze}{\frac{4}{3} \pi R_{\rm pot}^3}.
\end{equation}
The spin-orbit term has the form
\begin{equation}\label{eq:appr:pes:SO}
    V_{\rm s.o.}^{(\tau)} =
      \lambda_{\tau}
      \left(\frac{\hbar^2}{2m_{\rm nuc}c}\right)^2
      \frac{
               \nabla V^{(\tau)}_{1}
               \fvec{\sigma}
        \times \fvec{p}}{\hbar}
\end{equation}
where the interaction strength $\lambda_\tau$ is taken for each isospin $\tau$ as a linear function of the mass $A$~\cite{moeller1995nuclear} such that
\begin{equation}
    \lambda_\tau = k_\tau A + l_\tau \ ,
\end{equation}
and $k_\tau$ and $l_\tau$ are parameters. All the parameters involved in the calculation of our potential energy landscapes are listed in Appendix~\ref{app:FRLDM}.

Assuming that the particles of the compound system are independent, the many-body state can, therefore, be obtained as a Slater determinant of particles, the state of each particle being an eigenfunction of the Hamiltonian associated with the energy $e_i$ of the particle. 
Differentiating ourselves from past FRLDM work, the shell correction is calculated independently for each isospin by using the improved Strutinsky method presented in Ref.~\cite{kruppa1998calculation, tajima2010improved}.
This procedure removes spurious contribution from the continuum, which happens when the number of shells, $N_{0}$, in the Harmonic Oscillator basis used to estimate the shell-plus-pairing correction, is set too large, thus causing divergences in the calculation of the energy. 
The avoidance of such spurious contributions is the reason why the size of the particle basis was limited to $N_{0}=12$ in past work. In this work, we use $N_{0}=20$, which is sufficiently larger than $N_{0}=12$ while remaining computationally manageable. We have tested our Strutinksy procedure up to $N_{0}=30$ and found no anomalies. 

Pairing correlations are obtained using the Lipkin-Nogami approach with the seniority-pairing approximation on the Slater-determinant of particles used to estimate shell effects \cite{lipkin1960collective, nogami1964improved}. The relevant equations read
\begin{align}
    N_{\rm pair} &=
            L_{\rm min}
          + \sum_{k=L_{\rm min}}^{L_{\rm max}} v_k^2
           \label{eq:pes:LN0} \\
    \frac{2}{G} &=
            \sum_{k=L_{\rm min}}^{L_{\rm max}}
                \frac{1}{\sqrt{(\epsilon_k-\lambda)^2+\Delta^2}} \label{eq:pes:LN1}\\
    v_k^2 &=
      \frac{1}{2}
      \left[
          1
        - \frac{\epsilon_k - \lambda}%
               {\sqrt{(\epsilon_k - \lambda)^2+\Delta^2}}
      \right] \label{eq:pes:LN2}\\
    \epsilon_k &=
        e_k
      + (4\lambda_2-G) v_k^2 \label{eq:pes:LN3}\\
      \lambda_2 &=
        \frac{G}{4}
        \left\{
          \frac{\displaystyle
              \sum_{k=L_{\rm min}}^{L_{\rm max}}
                u_k^3 v_k
                \sum_{\substack{l=L_{\rm min}\\l\neq k}}^{L_{\rm max}}
                  u_l v_l^3
          }{\displaystyle
              \sum_{k=L_{\rm min}}^{L_{\rm max}}
                u_k^2 v_k^2
                \sum_{\substack{l=L_{\rm min}\\l\neq k}}^{L_{\rm max}}
                  u_l^2 v_l^2
            }
        \right\}. \label{eq:pes:LNlambda2}
\end{align}
This is a nonlinear system of $2N_{\rm v}+3$ equations, where $N_{\rm v}=L_{\rm max}-L_{\rm min}+1$ is the number of pairs in the valence space. The unknowns of this systems are the pairing gap $\Delta$, the Fermi energy $\lambda$, the number-fluctuation constant $\lambda_2$ and for $k=L_{\rm min} \dots L_{\rm max}$, $v_k$ are the occupation amplitudes and $\epsilon_k$ are the shifted single-particle energies. This system of equations is completely determined by the single-particle energies $e_k$, the number of paired levels $N_{\rm pair}$, the first and last levels $0\leq L_{\rm min} \leq L_{\rm max}$ in the valence space and the seniority-pairing strength, $G$. 
The latter is obtained using a method based on Ref.~\cite{madland1988new}. 
Specifically, to obtain an expression for $G$, we assume that the spacing between the energy levels is constant
\begin{equation}
    \epsilon_k - \tilde{\lambda} = \frac{k-N_{\rm pair}}{\tilde{\rho}}.
\end{equation}
In this expression, $\tilde{\lambda}$ is the smooth Fermi energy of the smoothed single-particle energy and $\tilde{g}$ is the smooth level density obtained with the Strutinsky method. The pairing gap $\Delta$ in~\eqref{eq:pes:LN2} is approximated by an effective pairing gap
\begin{equation}
    \bar{\Delta} =
        \begin{cases}
          r_{\rm mic} B_s/N^{1/3} & \text{for neutrons}\\
          r_{\rm mic} B_s/Z^{1/3} & \text{for protons}
        \end{cases}.
\end{equation}
We then substitute the sum of~\eqref{eq:pes:LN1} by an integral
\begin{equation}
    \sum_{k=L_{\rm min}}^{L_{\rm max}} f(\epsilon_k-\lambda)
      \approx \tilde{\rho} \int_{y_1}^{y_2} f(x) \diff{x}
\end{equation}
to obtain the following expression for $G$
\begin{equation}
    G =
      \frac{2}{\tilde{\rho}}
      \left[
        \operatorname{arsinh}\left(\frac{y_2}{\bar{\Delta}}\right)
      - \operatorname{arsinh}\left(\frac{y_1}{\bar{\Delta}}\right)
      \right]^{-1},
\end{equation}
where
\begin{align}
    y_1 &= \frac{L_{\rm min} - N_{\rm pair}-\frac{1}{2}}{\tilde{\rho}} \\
    y_2 &= \frac{L_{\rm max} - N_{\rm pair}+\frac{1}{2}}{\tilde{\rho}}.
\end{align}
Note that these expressions differ from the one obtained in~\cite{moeller1995nuclear,moller2016nuclear}, being shifted by $1/(2\tilde{\rho})$. The expression of the average pairing correlation plus quasi-particle energy $\tilde{E}_{\rm p.c.}$ is then obtained by inserting these quantities into Eq.~(110) of Ref.~\cite{moeller1995nuclear}.

We solve the Lipkin-Nogami equations~\eqref{eq:pes:LN0}-\eqref{eq:pes:LNlambda2} using a new method based on the analytical calculation of the full Lipkin-Nogami Jacobian and the action of its inverse on any vector coupled with a fifth-order numerical scheme. We set multiple starting points to avoid local minima and find that our method greatly enhances the success rate of the resolution of the Lipkin-Nogami equations. Thus, the time required to solve the Lipkin-Nogami equations using our method is faster than in older work. This means the calculation time for our procedure is now negligible compared to the calculation and diagonalization time of the Hamiltonian, thus allowing for larger shape families or potential energy surfaces to be explored in the future. This new method is described in detail in Appendix~\ref{app:LNsolve}. 

\subsection{Novel approach to strongly damped nuclear motion}
\label{subsec:appr:DPS}

Many different methods have been developed to calculate nascent fission fragment yields. Most of these methods seek to simplify the time-evolution of the complex nuclear motion subject to various assumptions, see Refs.~\cite{schmidt2018review, schunck2016microscopic} for recent reviews. 
One of the most successful methods pursued over the past ten years is the assumption of strongly damped nuclear dynamics in which the Smoluchowski equations reduce to Brownian-shape motion that may be described by a Metropolis random walk \cite{randrup2011brownian, randrup2011fission}. 
This method has been used in a large range of fission reactions using Markov Chain Monte Carlo (MCMC) sampling~\cite{randrup2013energy, moller2012calculated, moller2014fission, moller2015calculated, mumpower2020primary}. 
The primary drawback to this method is the long calculation time (due to assimilation of enough trajectories for sufficient statistics) required for estimating fission properties of a fissioning system whose energy is slightly above the fission barrier. 
In what follows, we briefly review the construction of the finite-temperature PES and the MCMC sampling method. 
We then present our new deterministic algorithm (DPS) for attaining scissioned configurations under the assumption of strongly damped motion. 
In this work, we show that our deterministic algorithm is completely equivalent to the standard one, used in Ref.~\cite{mumpower2020primary}. 
Our new algorithm affords no statistical error associated with its results and allows rapid simulation of fission reactions with an excitation energy close to the energy of the fission barrier. 

\subsubsection{The standard approach (Monte Carlo)}
\label{sec:appr:DPS:former}

Following Ref.~\cite{mumpower2020primary}, the finite-temperature potential energy $U(\MQS{\chi})$ for each points of the PES is obtained through the insertion of a suppression factor $\mathcal{S}[E^*(\MQS{\chi})]$ in Eq.~\eqref{eq:appr:pes:Etot},
\begin{align}
  \label{eq:appr:DPS:dampingE}
  U(\MQS{\chi}) &=
      E_{\rm mac}(\MQS{\chi})
    + \mathcal{S}[E^*(\MQS{\chi})]
        \Delta E_{\rm s+p}(\MQS{\chi}) \\
  \Delta E_{\rm s+p}(\MQS{\chi}) &=
        \Delta E_{\rm shell}(\MQS{\chi})
      + \Delta E_{\rm pair.}(\MQS{\chi}) \\
  \label{eq:appr:DPS:dampingS}
  \mathcal{S}[E^*] &=
    \frac{1+\operatorname{exp}(-E_1/E_0)}%
         {1+\operatorname{exp}((E^*-E_1)/E_0)} \ ,
\end{align}
where $E^*(\MQS{\chi}) = E^* - E(\MQS{\chi})$ is the local excitation energy and $E_0$ and $E_1$ are two parameters that control the shell damping term, as in Ref.~\cite{randrup2013energy}. 
A discrete random walk is then used on the finite-temperature PES $U(\MQS{\chi})$ to ascertain the scissioned configurations for the given incident energy, $E^*$. 

A single trajectory in the standard MCMC random walk follows the evolution of a path through the PES represented by a sequence of contiguous steps.
A trajectory begins at an initial shape, $\MQS{\chi}_{\rm init}$, and stops when an ending configuration, $\MQS{\chi}_{\rm end}$, is reached.
Ending configurations are taken to be the scissioned shapes (e.g.~ the radius of its neck, if present, is lower than a given parameter $r_{\rm neck}$). 

In the MCMC procedure, a step is a move between a shape $\MQS{\chi}$ and one of its $D$ neighbors $\MQS{\chi}_k$ ($k=1,\dots,D$) and is denoted $\MQS{\chi} \Rightarrow \MQS{\chi}_k$. 
The determination of one step proceeds via intermediate moves along a Markov chain, which we call transitions. 
The chain can be represented by a directed graph with a set of nodes associated with each lattice site of the PES. 
A particular node in the Markov chain is denoted by $\chi_0$, which implicitly depends on the shape $\MQS{\chi}$. 
Upon reaching this node, it triggers a step to lattice site $\MQS{\chi}$. 

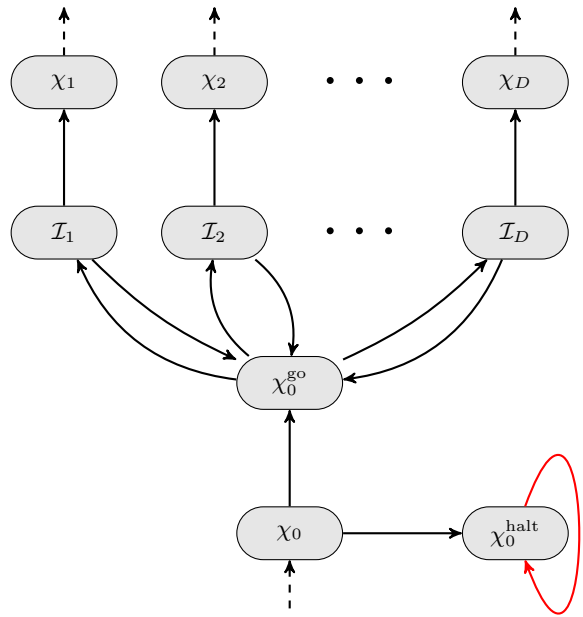
\begin{figure}[ht]
  \centering
  \tikzset{every loop/.style={min distance=10mm,in=-70,out=70,looseness=10}}
  \begin{tikzpicture}[->,>=stealth',auto,node distance=3cm,thick,RWnode/.style={draw,thin,rectangle,rounded corners=10,draw,align=center,minimum height=20, minimum width=40, fill=gray!20}]
    \node[RWnode] (init) at ( 0, 0) {$\chi_0$};
    \node[RWnode] (go)   at ( 0, 2) {$\chi_0^{\rm go}$};
    \node[RWnode] (halt) at ( 3, 0) {$\chi_0^{\rm halt}$};
    \node[RWnode] (I1) at (-3, 4) {$\mathcal{I}_1$};
    \node[RWnode] (I2) at (-1, 4) {$\mathcal{I}_2$};
    \node[RWnode] (ID) at ( 3, 4) {$\mathcal{I}_D$};
    \node[RWnode] (X1) at (-3, 6) {$\chi_1$};
    \node[RWnode] (X2) at (-1, 6) {$\chi_2$};
    \node[RWnode] (XD) at ( 3, 6) {$\chi_D$};
  
    \path [every node/.style={font=\sffamily\small}]
        ( 0,-1) edge[dashed] node {} (init)
        (init)  edge node {} (go)
        (init)  edge node {} (halt)
        (go)    edge [bend left]  node {} (I1)
        (I1)    edge [bend left=-10] node {} (go)
        (go)    edge [bend left]  node {} (I2)
        (I2)    edge [bend left] node {} (go)
        (go)    edge [bend left=-10]  node {} (ID)
        (ID)    edge [bend left] node {} (go)
        (I1)    edge node {} (X1)
        (I2)    edge node {} (X2)
        (ID)    edge node {} (XD)
        (X1)    edge[dashed] node {} (-3, 7)
        (X2)    edge[dashed] node {} (-1, 7)
        (XD)    edge[dashed] node {} ( 3, 7)
        (I2)    -- node[auto=false] {\Huge$\cdots$} (ID)
        (X2)    -- node[auto=false] {\Huge$\cdots$} (XD)
        (halt)  edge[loop above,red!100] node {} (halt);
  \end{tikzpicture}
  \caption{
    Graph of the Markov chain transitions determining one step the standard MCMC random walk method (without the red transition, where $\chi_0^{\rm halt}$ is a final node of the Markov chain). In our novel algorithm, the red looping transition is added to transform $\chi_0^{\rm halt}$ into an absorbing node so that an execution of the Markov chain cannot terminate.\label{fig:appr:DPS:RW}
  }
\end{figure}

The sub-graph of the Markov chain depicting a single step in the lattice is presented in \figref{fig:appr:DPS:RW}. A single step in a trajectory can be decomposed into the procedure below. 
\begin{enumerate}
    \item Start at a shape $\MQS{\chi}$ (node $\chi_0$). Set the transition probabilities accordingly, 
        \begin{align}
            \prob{\chi_0\rightarrow\chi_0^{\rm go}} &=
            \begin{cases}
              0 & \text{$\MQS{\chi}$ is scissioned} \\
              1 & \text{otherwise}
            \end{cases} \\
            \prob{\chi_0\rightarrow\chi_0^{\rm halt}} &= 1 - \prob{\chi_0\rightarrow\chi_0^{\rm go}} \ .
        \end{align}
        In this case, make the trivial transition, $\chi_0\rightarrow\chi_0^{\rm go}$ or $\chi_0\rightarrow\chi_0^{\rm halt}$. Stop the calculation if $\chi_0^{\rm halt}$ is reached. 
    \item Randomly choose one of the $D$ neighbors $\MQS{\chi}_k$ ($k=1,\dots,D$) using a discrete uniform distribution (transition $\chi_0^{\rm go}\rightarrow\mathcal{I}_k$).
    \item Compare the finite-temperature potential energies $\Delta U = U(\MQS{\chi}_k) - U(\MQS{\chi})$. Then, set
          \begin{equation}
            \prob{\mathcal{I}_k\rightarrow\chi_k} =
              \begin{cases}
                \operatorname{exp}
                \left(
                  -\frac{\Delta U}{T}
                \right) & \text{if $\Delta U > 0$} \\
                \qquad 1 & \text{otherwise.}
                \end{cases}
          \end{equation}
          where $T=[E^*(\MQS{\chi})/a]^{1/2}$ is the local effective temperature and $a$ is the level-density parameter taken to be $a=A/8$~\cite{mumpower2020primary}.
    \item Make the transition $\mathcal{I}_k\rightarrow\chi_k$ with probability $\prob{\mathcal{I}_k\rightarrow\chi_k}$. If the transition occurs, go to the node $\chi_0(\MQS{\chi}_k)$ of the neighboring shape $\MQS{\chi}_k$ ($\chi_k\rightarrow\chi_0(\MQS{\chi}_k)$) and perform the step $\MQS{\chi}\Rightarrow\MQS{\chi}_k$ in the lattice. Otherwise, go back to $\chi_0^{\rm go}$ (transition $\mathcal{I}_k\rightarrow\chi_0^{\rm go}$) and repeat procedure from step 2.
\end{enumerate}
This procedure defines all of the intermediate transitions that determine one step in the lattice, differing slightly from the standard interpretation, since it is impossible to step from a lattice site to itself. 
The standard interpretation of the MCMC algorithm can be recovered by also triggering a step, $\MQS{\chi}\Rightarrow\MQS{\chi}$, any time a transition $\mathcal{I}_k\rightarrow\chi_0^{\rm go}$ is made. 
Since we are only interested in determining the distribution of ending configurations, this subtle point does not impact outcomes, but is necessary for the transformation required by the novel algorithm presented in the next section.

In previous work, the combination of trajectories corresponding to realizations of the full Markov chain are simulated to produce the set of scissioned configurations. 
Many trajectories have to be calculated to reduce the statistical uncertainty and obtain a reasonable estimation of the probability density function for the scissioned configurations. 

\subsubsection{Deterministic algorithm (without absorption)}

In contrast to the above approach, we propose to directly determine the evolution of the probability density function for the configurations in the lattice according to the number of steps in a trajectory associated with the MCMC random walk. 
This transformation is analogous to the determination of the Fokker-Plank equations associated with the Langevin equations. 
In what follows, we use $\chi\leadsto\varphi$ to indicate the set of all possible paths between $\chi$ and $\varphi$ in the sub-graph of the Markov chain associated with one lattice site, plus possibly the sites $\chi_0(\MQS{\chi})'$ of the neighboring sites $\MQS{\chi}'$. 

We first determine the probabilities associated with any step from a lattice site $\MQS{\chi}$ to a neighboring site $\MQS{\chi}'$.
This probability, $\prob{\MQS{\chi}\Rightarrow\MQS{\chi}'}$, is equal to the probability $\prob{\chi_0\leadsto\chi_0(\MQS{\chi}')}$ to reach the node $\chi_0(\MQS{\chi}')$ from the node $\chi_0$ in the Markov chain.
Any path from $\chi_0$ to $\chi_0(\MQS{\chi}')$ can be decomposed into three independent parts: (i) $\chi_0\rightarrow\chi_0^{\rm go}$, (ii) $\chi_0^{\rm go}\rightarrow\dots\rightarrow\chi_k$ (iii) $\chi_k\rightarrow\chi_0(\MQS{\chi}')$.
Using the so-called \textit{memoryless} property of the Markov chain and the triviality of the last transition $\chi_k\rightarrow\chi_0(\MQS{\chi}')$ (associated with probability 1), we have
\begin{equation}\label{eq:appr:DPS:firstmemless}
    \prob{\MQS{\chi}\Rightarrow\MQS{\chi}'} =
      \prob{\chi_0\rightarrow\chi_0^{\rm go}}
      \times\prob{\chi_0^{\rm go}\leadsto\chi_k}\ .
\end{equation}
We focus now on the determination of the probability $\prob{\chi_0^{\rm go}\leadsto\chi_k}$ to reach the node $\chi_k$ from $\chi_0^{\rm go}$ through any path of the Markov chain of \figref{fig:appr:DPS:RW}, from the node $\chi_0^{\rm go}$. To do so, let's notice that any path $\mathcal{P}$ between the nodes $\chi_0^{\rm go}$ and $\chi_k$ can be decomposed into two parts. First, $\mathcal{P}$ starts with a loop from $\chi_0^{\rm go}$ to itself containing an even number of transitions, denoted $\mathcal{L}(\mathcal{P})$. Second, it ends with the two transitions $\chi_0^{\rm go}\rightarrow\mathcal{I}_k\rightarrow\chi_k$. Using the memoryless property of Markov chains, this translates into the equation, 
\begin{equation}
    \prob{\mathcal{P}} = \prob{\mathcal{L}(\mathcal{P})}\frac{\prob{\mathcal{I}_k\rightarrow\chi_k}}{D}\ .
\end{equation}
Recall, $D$ is the number of neighboring lattice sites. 
The probability $\prob{\chi_0^{\rm go}\leadsto\chi_k}$ to reach the node $\chi_k$ from $\chi_0^{\rm go}$ is the sum of the probabilities of all path $\mathcal{P}$ from $\chi_0^{\rm go}$ to $\chi_k$, 
\begin{align}
    \label{eq:appr:DPS:probdecompo}
    \prob{\chi_0^{\rm go}\leadsto\chi_k} &=
        C \times
        \frac{\prob{\mathcal{I}_k\rightarrow\chi_k}}{D}\ , \\
    C &= \sum_{\mathcal{P}}\prob{\mathcal{L}(\mathcal{P})}\ .
\end{align}
Any path starting at $\chi_0^{\rm go}$ has to pass through exactly one of the nodes $\chi_k$. Therefore, the sum over $k=1,\dots,D$ of Eq.~\eqref{eq:appr:DPS:probdecompo} adds up to one. 
This leads to
\begin{equation}\label{eq:appr:DPS:Cexpr}
    C = \left[\sum_{k=1}^{D}\frac{\prob{\mathcal{I}_k\rightarrow\chi_k}}{D}\right]^{-1} \ .
\end{equation}
Alternatively, the expression of $C$ can be obtained by using again the memoryless transitions property to easily show that the probability $\prob{\chi_0^{\rm go}\stackrel{2N}{\leadsto} \chi_0^{\rm go}}$ to reach $\chi_0^{\rm go}$ from itself after $2N$ transitions is
\begin{equation}\label{eq:appr:DPS:00exp}
    \prob{\chi_0^{\rm go}\stackrel{2N}{\leadsto} \chi_0^{\rm go}} =
      \left[
        \sum_{k=1}^D \frac{\prob{\mathcal{I}_k\rightarrow\chi_0^{\rm go}}}{D}
      \right]^N \ .
\end{equation}
Since it is impossible to reach $\chi_0^{\rm go}$ in an odd number of transitions, we have directly
\begin{equation}
    \prob{\chi_0^{\rm go}\stackrel{2N+1}{\leadsto} \chi_0^{\rm go}} = 0 \ .
\end{equation}
The coefficient $C$ is the sum of all the closed loops passing through $\chi_0^{\rm go}$. Consequently,
\begin{equation}
    C = \sum_{N=0}^{\infty} \left[
        \sum_{k=1}^D \frac{\prob{\mathcal{I}_k\rightarrow\chi_0^{\rm go}}}{D}
      \right]^N\ .
\end{equation}
Eq.~\eqref{eq:appr:DPS:Cexpr} can directly be obtained by recognizing the Taylor series of the function $x\mapsto 1/(1-x)$ and using the relation $\prob{\mathcal{I}_k\rightarrow\chi_0^{\rm go}}+\prob{\mathcal{I}_k\rightarrow\chi_k}=1$.

The probability $\prob{\chi_0^{\rm go}\leadsto\chi_k}$ in the Markov chain of~\figref{fig:appr:DPS:RW} is directly obtained by inserting Eq.~\eqref{eq:appr:DPS:Cexpr} into Eq.~\eqref{eq:appr:DPS:probdecompo}, 
\begin{multline}\label{eq:model:pchichi}
    \prob{\chi_0^{\rm go}\leadsto\chi_k} =
      \left[\sum_{k=1}^{D}\frac{\prob{\mathcal{I}_k\rightarrow\chi_k}}{D}\right]^{-1} \\
      \times \frac{\prob{\mathcal{I}_k\rightarrow\chi_k}}{D} \ .
\end{multline}
The corresponding transition probability, $\prob{\MQS{\chi}\Rightarrow\MQS{\chi}'}$, from a lattice site $\MQS{\chi}$ to a neighboring site $\MQS{\chi}'$ is proportional to $\prob{\mathcal{I}_k\rightarrow\chi_k}$. Our derivation shows that in the standard random walk method, there exists a parameter that correspond to an energy threshold, $\Delta U_{\rm thresh.}=0$ MeV, on the energy difference, $\Delta U = U(\MQS{\chi}_k)-U(\MQS{\chi}_0)$, below which a transition from $\mathcal{I}_k$ to $\chi_k$ is certain to happen (with probability 1). 

With the probability to reach $\chi_k$ from $\chi_0^{\rm go}$ well defined, we now seek to calculate the probability distribution, $p^{(n)}(\MQS{\chi})$, after a fixed number of random walk steps, $n$. 
Neglecting any halting condition, this quantity can be obtained recursively at the $(n+1)$-th iteration from the distribution $p^{(n)}(\MQS{\chi})$ using the relation
\begin{equation}\label{eq:model:pdfrw}
    p^{(n+1)}(\MQS{\chi}) =
      \sum_{\MQS{\chi}'}
        p^{(n)}(\MQS{\chi}')
        \mathcal{T}(\MQS{\chi}', \MQS{\chi}) \ ,
\end{equation}
where $\mathcal{T}(\MQS{\chi}, \MQS{\chi}')=\prob{\chi_0^{\rm go}(\MQS{\chi})\leadsto\chi_k(\MQS{\chi})}$. 
Because the halting condition is neglected, $\mathcal{T}(\MQS{\chi}, \MQS{\chi}')$ is obtained from $ \prob{\MQS{\chi}\Rightarrow\MQS{\chi}'}$ by setting $\prob{\chi_0\rightarrow\chi_0^{\rm go}}=1$ in~\eqref{eq:appr:DPS:firstmemless}. 
From this definition it is clear that the probability distribution after $n$ steps is thus completely determined by the choice of the initial probability distribution $p^{(0)}(\MQS{\chi})$. 

\subsubsection{Generalized deterministic algorithm (with absorption)}

The deterministic algorithm consists of evaluating $p^{(n)}(\MQS{\chi})$ iteratively through the use of Eq.~\eqref{eq:model:pdfrw}. 
Rather than directly using Eq.~\eqref{eq:model:pdfrw} to compute our scission probability distribution, we may generalize our procedure to allow for a distribution of ending configurations. 
This amounts to adding an absorption mechanism on top of the nuclear PES. 

It is important to recognize that each MCMC trajectory can have a completely different number of steps, while we aim at determining the evolution of the probability density of all the trajectories at the same time (requiring an equal number of steps). 
Therefore, we add one site $\MQS{\chi}^{\rm halt}$ for each site $\MQS{\chi}$ in our lattice. 
Each trajectory from an initial site and reaching an ending configuration $\MQS{\chi}_{\rm end}$ in the standard MCMC algorithm can then be equalized in length by adding the steps $\MQS{\chi}_{\rm end}\Rightarrow\MQS{\chi}_{\rm end}^{\rm halt}\Rightarrow\cdots\Rightarrow\MQS{\chi}_{\rm end}^{\rm halt}$ as many times as necessary, represented as node $\chi_0^{\rm halt}$ in \figref{fig:appr:DPS:RW}. 
Node $\chi_0^{\rm halt}$ can be considered an absorbing node and we interpret this action as reaching of the (new) lattice site $\MQS{\chi}_{\rm end}^{\rm halt}$ through the step $\MQS{\chi}_{\rm end}\Rightarrow\MQS{\chi}_{\rm end}^{\rm halt}$. 
Since all the sites $\MQS{\chi}^{\rm halt}$ are absorbing and only accessible from $\MQS{\chi}$, we have, for any two neighboring lattice sites $\MQS{\chi}_1$ and $\MQS{\chi}_2$,
\begin{align}
  \prob{\MQS{\chi}_1^{\rm halt}\Rightarrow\MQS{\chi}_2}&=0\\
  \prob{\MQS{\chi}_1^{\rm halt}\Rightarrow\MQS{\chi}_2^{\rm halt}} &= \delta_{\MQS{\chi}_1, \MQS{\chi}_2} \\
  \prob{\MQS{\chi}_1\Rightarrow\MQS{\chi}_2^{\rm halt}} &= \delta_{\MQS{\chi}_1, \MQS{\chi}_2}\mathcal{A}(\MQS{\chi}_1)\ ,
\end{align}
where
\begin{equation}
    \mathcal{A}(\MQS{\chi})=\prob{\chi_0\rightarrow\chi_0^{\rm halt}}
\end{equation}
is the absorption field. 

We let $\bar{a}^{(n)}(\MQS{\chi})$ denote the density of probability at the lattice site $\MQS{\chi}^{\rm halt}$ after $n$ random walk steps in the PES. 
We now proceed to determine the evolution of the density of probability. 
Invoking the memoryless property of Markov chains one last time, we have
\begin{equation}
    \prob{\MQS{\chi}\Rightarrow\MQS{\chi}'} = \left[1-\mathcal{A}(\MQS{\chi})\right]\mathcal{T}(\MQS{\chi}, \MQS{\chi}')\ .
\end{equation}
Thus, the full recurrence relation on the $p^{(n)}(\MQS{\chi})$ and the $\bar{a}^{(n)}(\MQS{\chi})$ now reads
\begin{align}
    \label{eq:model:probevolv}
    p^{(n+1)}(\MQS{\chi}) &=
      \sum_{\MQS{\chi}'}
        p^{(n)}(\MQS{\chi}')
        \left[1-\mathcal{A}(\MQS{\chi}')\right]\mathcal{T}(\MQS{\chi}', \MQS{\chi}) \\
    \label{eq:model:probevolvabso}
    \bar{a}^{(n+1)}(\MQS{\chi}) &=
      \bar{a}^{(n)}(\MQS{\chi})
    + p^{(n)}(\MQS{\chi})
      \mathcal{A}(\MQS{\chi})\ ,
\end{align}
where the normalization of the distributions $p^{(n)}(\MQS{\chi})$ and $\bar{a}^{(n)}(\MQS{\chi})$ now reads
\begin{equation}
    \sum_{\MQS{\chi}} [p^{(n)}(\MQS{\chi}) + \bar{a}^{(n)}(\MQS{\chi})] = 1\ .
\end{equation}
Finally, we introduce the probability distribution, $a^{(n)}(\MQS{\chi})$, to reach the configuration $\MQS{\chi}$ knowing that it is absorbed,
\begin{equation}
    \label{eq:model:absotot}
    a^{(n)}(\MQS{\chi}) = \frac{\bar{a}^{(n)}(\MQS{\chi})}{\sum_{\MQS{\chi}'}\bar{a}^{(n)}(\MQS{\chi}')}\ .
\end{equation}
Our deterministic algorithm thus proceeds by iteratively evaluating Eqs.~\eqref{eq:model:probevolv} and~\eqref{eq:model:probevolvabso}. 
After $n$ steps, the probability of reaching a scissioned configurations (on an absorbed node) is finally calculated with \eqref{eq:model:absotot}. 

The above absorption mechanism used to simulate the stopping of the calculation in the standard MCMC random walk is analogous to the procedure used in Ref.~\cite{berger1991time} in case of the absorption of the TDGCM+GOA wavefunction for scissioned configurations. 
In this first work, we set the absorption field, $\mathcal{A}(\MQS{\chi})$, to 100\% for scissioned configurations ($\mathcal{A}(\MQS{\chi})=1$) and zero otherwise ($\mathcal{A}(\MQS{\chi})=0$). 
This binary selection ensures that our algorithm is formally equivalent to the standard random walk approach up to the statistical error of the standard algorithm as well as a possible numerical error due to the finite representation of real numbers in computers. 
Comparisons between the results obtained using the new algorithm with the standard approach (that explicitly samples many trajectories) are presented in Section~\ref{sec:appli}. 
We note that the absorption field can in principle be configured in any number of ways, including a distribution for various neck sizes, instead of the sharp function used in this work. 
We plan to study such possibilities in future work. 

\subsubsection{Convergence criteria of the new algorithm}

Our novel method also affords the calculation of a convergence criteria. We can use this quantity as an estimate of the statistical uncertainty on calculated fission yields. To this end, we assume that the 1-distance (the distance associated with the 1-norm $||.||_1$) between two distributions $a^{(n)}$ and $a^{(n+j)}$, defined as
\begin{equation}\label{eq:appr:cc}
    \Delta_{(j)}^{(n)} =
      \sum_{\MQS{\chi}}
        \left|
            a^{(n+j)}(\MQS{\chi})
          - a^{(n  )}(\MQS{\chi})
        \right|,
\end{equation}
is an inverse-quadratic function
\begin{equation}\label{eq:appr:DPS:cvgapprox}
  \Delta_{(j)}^{(n)} \approx \frac{1}{[c(j)n+d(j)]^2} \ ,
\end{equation}
where $j$ is an integer, $c(j)$ and $d(j)$ are two real parameters that are ultimately obtained using a fit procedure. 
The validity of this assumption is discussed in \secref{sec:appli} for $j=1000$.
The convergence error at a step $n$ can be introduced as the distance between the distribution at step $n$ and the distribution at infinity
\begin{equation}
    \varepsilon^{(n)} = \Delta_{(\infty)}^{(n)}.
\end{equation}
Using the subadditivity of the 1-distance (commonly referred as the triangle inequality), we can determine the convergence criteria,
\begin{equation}\label{eq:appr:DPS:trileq}
    \varepsilon^{(n)} \leq \sum_{k=0}^{\infty} \Delta_{(j)}^{(n+kj)}\ .
\end{equation}
Inserting~\eqref{eq:appr:DPS:cvgapprox} into~\eqref{eq:appr:DPS:trileq} gives
\begin{equation}
    \varepsilon^{(n)} \leq \sum_{k=0}^{\infty}\frac{1}{\left[(jc(j))k+(c(j)n+d(j))\right]^2}\ .
\end{equation}
A closed form of the right-hand side of this expression can be obtained by resorting to the trigamma function defined as
\begin{equation}
    \phi^{(1)}(z) =
      \frac{\diff{}^2}{\diff{z^2}}\operatorname{ln}\Gamma(z)\ ,
\end{equation}
that satisfies
\begin{equation}
    \phi^{(1)}(z) = \sum_{k=0}^{\infty}\frac{1}{(k+z)^2}\ .
\end{equation}
Employing these properties of the trigamma function, we finally obtain
\begin{equation}
    \varepsilon^{(n)} \leq \varepsilon^{(n,j)}_{\rm eff.}\ ,
\end{equation}
where
\begin{equation}\label{eq:converg:upper}
  \varepsilon^{(n,j)}_{\rm eff.} =
      \frac{1}{[jc(j)]^2}
        \phi^{(1)}
        \left(
          \frac{c(j)n+d(j)}{jc(j)}
        \right) \ .
\end{equation}
We take $\varepsilon^{(n,j)}_{\rm eff.}$ as our convergence criteria as it is an upper limit on the error $\varepsilon^{(n)}$ after $n$ iterations.

\subsection{Mass and charge yields of fission fragments}
\label{subsec:appr:frag}

Most of the current mac-mic models used to estimate the probability distribution associated with the fission fragment properties before prompt emissions resort to obtaining the mass, $Y(\Afrag)$, or charge, $Y(\Zfrag)$, yields separately using the relevant macroscopic shape parameter, e.g., the procedure of Ref.~\cite{moller2014fission}. 
While this method has been highly successful, see e.g.~\cite{randrup2011fission, randrup2011brownian}, it does not provide a means to calculate the full fragment yield, $Y(\Zfrag, \Afrag)$ or equivalently, $Y(\Zfrag, \Nfrag)$. 
In the past, the full mass and charge yields have been obtained through the direct analysis of systematics on known experimental data \cite{caamano2011evidence}, thus with low predictive power albeit high-quality data, or using the Wahl systematics~\cite{wahl2002systematics} e.g., in Ref.~\cite{mumpower2020primary}, which introduces a free parameter $\sigma_Z$ that controls the dispersion in charge of the isobaric yields, and presumes the Unchanged Charge Distribution (UCD) assumption that relies on the ratio $\eta \equiv \frac{Z}{A} = \frac{\Zfrag}{\Afrag}$. 
Another method, presented in Ref.~\cite{moller2015method}, aims at obtaining the full mass and charge yields but relies on the addition of a sixth macroscopic shape parameter in the PES. Without adding specific parameters, none of these past approaches can predict the fission fragments' charge polarization, which is the experimentally observed deviation from the UCD assumption.

In the following, we construct an approach to predict the full probability distribution of the fission fragment mass \textit{and} charge $Y(\Zfrag, \Afrag)$ directly from the quantum mechanical wavefunctions.
This method directly probes the structure of the nascent fragments, enabling the description of fine-structural effects such as odd-even staggering and charge polarization observed in experimental data. 
As it will be shown in Sec.~\ref{subsec:appli:results}, our projection technique is able to reproduce these phenomenon without additional parameters. 

We only consider here the nascent fragments in binary fission. Therefore, $Y(\Zfrag, \Afrag)$ can be decomposed according to the probability distribution, $\YL(\Zfrag, \Afrag)$, (normalized to 1) associated with the mass and charge of only the left fragment,
\begin{equation}\label{eq:appr:frag:YLtoY}
    Y(\Zfrag, \Afrag) = \YL(\Zfrag, \Afrag)
                           + \YL(Z-\Zfrag, A-\Afrag) \ .
\end{equation}
The law of total probability enables the decomposition of the probability distribution of the number of particles in the left fragment $\YL(\Zfrag, \Afrag)$ before prompt particle emission as follows,
\begin{multline}\label{eq:appr:frag:StoYL}
  \YL(\Zfrag, \Afrag) =
    \int
      \prob[\MQS{\chi}]%
           {(Z, A)_{\rm L} = (\Zfrag, \Afrag)} \\
      \times a^{(\infty)}(\MQS{\chi})
      \diff{\MQS{\chi}} \ ,
\end{multline}
where the integral iterates over all the scission shapes parameterized by $\MQS{\chi}$, $a^{(\infty)}(\MQS{\chi})$ is the limit for $n\to\infty$ of the sequence of distributions introduced in Eq.~\eqref{eq:model:absotot} and $\prob[\MQS{\chi}]{(Z,A)_{\rm L} = (\Zfrag, \Afrag)}$ is the probability associated with a left fragment of mass $\Afrag$ and charge $\Zfrag$ when the fissioning system is in the shape $\MQS{\chi}$.

The probability $\prob[\MQS{\chi}]{(Z, A)_{\rm L} = (\Zfrag, \Afrag)}$ is extracted from the microscopic state calculated to estimate the shell+pairing correction for each coordinate $\MQS{\chi}$ of the PES after projection on the good mass and charge of the total fissioning system. Because all these states preserve the isospin, we can further decompose
\begin{multline}\label{eq:appr:frag:isospindecompo}
    \prob[\MQS{\chi}]{(Z, A)_{\rm L} = (\Zfrag, \Afrag)} =
      \prob[\MQS{\chi}]{N_{\rm L} = \Nfrag} \\
      \times\prob[\MQS{\chi}]{Z_{\rm L} = \Zfrag},
\end{multline}
where $\Nfrag$ is the number of neutrons in the left fragment. Both factors in the right-hand side of Eq.~\eqref{eq:appr:frag:isospindecompo} are calculated through
the particle-number projection-based technique on the fragments mass and charge developed first in Ref.~\cite{simenel2010particle} in the context of time-dependent mean-field calculations for transfer reactions. This technique was first applied to fission in Ref.~\cite{scamps2015superfluid} and adapted to the case of static mean-field calculations in Ref.~\cite{verriere2019number}. When applied to fission, this technique gives the probabilities associated with the number of nascent fragment neutrons ($X=N$, $\Xfrag = \Nfrag$) and protons ($X=Z$, $\Xfrag = \Zfrag$) using,
\begin{equation}
  \prob[\MQS{\chi}]{X_{\rm L} = \Xfrag} =
    \frac{
      \braket[\hat{P}^{(L)}_{\Xfrag}\hat{P}_X]{\Phi(\MQS{\chi})}{\Phi(\MQS{\chi})}
    }{
      \braket[\hat{P}_X]{\Phi(\MQS{\chi})}{\Phi(\MQS{\chi})}
    },
\end{equation}
or equivalently
\begin{equation}\label{eq:appr:frag:doubleP2}
  \prob[\MQS{\chi}]{X_{\rm L} = \Xfrag} =
    \frac{
      \braket[\hat{P}^{(L)}_{\Xfrag}\hat{P}^{(R)}_X]{\Phi(\MQS{\chi})}{\Phi(\MQS{\chi})}
    }{
      \braket[\hat{P}_X]{\Phi(\MQS{\chi})}{\Phi(\MQS{\chi})}
    }.
\end{equation}
A double projection is required, where $\hat{P}_{X=N,Z}$ is the operator restoring the good number of particle in the total system while $\hat{P}^{(L)}_{\Xfrag}$ is an operator projecting on $\Xfrag$ particles in the left fragment. The definition of the latter relies on the position of the neck along the symmetry axis. We define this quantity, in the standard way, as the position of the minimum of the local one-body density of the microscopic state in $\MQS{\chi}$ between the two pre-fragments~\cite{younes2009microscopic, regnier2018felix, perez2017axially}. 
The projection-based method to calculate the fragment distribution is already known to describe the odd-even staggering of the charge distribution of the fragments in the case of time-dependent mean-field methods~\cite{scamps2015superfluid}. It is also able to give a non-zero probability for the existence of fragments with an odd-number of particles~\cite{verriere2019number}. A simple example to illustrate how a time-even state can describe fragments with an odd number of particles is given in Appendix~\ref{app:nonzeroodd}.

\section{Application to 233,235-U(n,f)}
\label{sec:appli}

We illustrate our model improvements in what follows by showcasing the well-known neutron-induced fission of two isotopes of Uranium, \textsuperscript{233,235}U(n,f). We also prove that our implementation can reproduce the results of the discrete random-walk method used in past work.

In Refs.~\cite{moeller1995nuclear, moller2004five, moller2009heavy, moller2012calculated, moller2014fission, moller2015calculated, moller2016nuclear}, the authors implemented shell-plus-pairing corrections through the resolution of the Schr\"odinger equation using an axial harmonic oscillator basis with only $N_{\rm sh}=12$ shells. The limitation of this previous approach artificially introduces spurious contributions of the continuum \cite{tajima2010improved}. The use of the improved Strutinsky method~\cite{kruppa1998calculation} allows us to remove these contributions and thus use instead $N_{\rm sh}=20$ shells without any energy truncation. In addition, we optimize the oscillator scaling factor $b_0$ and deformation $q$ for each point of the PES using a variational principle such that the mean-field-plus-pairing energy of the microscopic state is minimized. 

The order of the Strutinsky method is $p=8$, and the corresponding range is 
\begin{equation}
  \gamma = C_{\rm sr} \frac{C_{\rm cur}}{A^{1/3}} B_{\rm s},
\end{equation}
where the relative surface energy $B_{\rm s}(\MQS{\chi})$ is the ratio of the nucleus surface at shape $\MQS{\chi}$ with the surface of the same nucleus at spherical shape. We have used the same parameters as in Ref.~\cite{moeller1995nuclear}, listed in the following table. 
\begin{table}[ht]
    \centering
    \begin{ruledtabular}
    \begin{tabular}{ccc}
      Parameter     & Value & Unit \\
      \hline
      $C_{\rm sr} $ & 1.0       & MeV \\
      $C_{\rm cur}$ & 41        & MeV
    \end{tabular}
    \end{ruledtabular}
    \caption{Microscopic parameters associated with the Strutinsky correction.}
    \label{tab:appli:param:FRLDM2002-strut}
\end{table}

The pairing correction is obtained using the Lipkin-Nogami method. 
We have solved the Lipkin-Nogami (LN) for each point of the PES using a pairing window of $\pm$ 5 MeV around the Fermi surface. The only remaining parameter is the LN effective-interaction pairing gap constant $r_{\rm mic}=$ 3.2 MeV. The full LN equations are often numerically solved by splitting them into two or more subsets of equations, solved separately at each iteration. It adds overhead in the resolution time and can also lead to spurious divergences. Instead, we have developed a new iterative method to solve these equations, presented in Appendix~\ref{app:LNsolve}.

The fragment probabilities at each scissioned configuration $\MQS{\chi}$ are calculated  using Eq.~\eqref{eq:appr:frag:doubleP2}.
The double-projection on the numerator is calculated using the Pfaffian technique presented in Refs.~\cite{bertsch2012symmetry, robledo2009sign}.
The determination of the integrals over the gauge angles are determined through the Fomenko discretization method~\cite{fomenko1970projection}. 
The number of integration nodes $n_{\rm Fom.}$ is determined for each isospin of each scissioned configuration starting with $n_{\rm Fom.}=30$. 
We then check if the inequality (where $X$ has to be replaced by $N$ or $Z$ according to the isospin)
\begin{equation}
  \label{eq:appli:param:fom-nnodes-cond}
  \prob[\MQS{\chi}]{X_{\rm L} = \Xfrag} <
    1.0\times 10^{-6}
\end{equation}
is satisfied for all $\Xfrag = \lfloor X_{\rm mean}-n_{\rm Fom.}/2\rfloor+\Delta X$ and $\Xfrag = \lceil X_{\rm mean}+n_{\rm Fom.}/2\rceil-\Delta X$ for $\Delta X=0,1,2$.
If the relation~\eqref{eq:appli:param:fom-nnodes-cond} is not satisfied, we add 10 to $n_{\rm Fom.}$ and test again our criteria up to finding a suitable value of $n_{\rm Fom.}$.

The PES at a given excitation energy is obtained through the finite temperature method of Ref.~\cite{randrup2013energy} where the damping of the shell-plus-pairing correction as in Eq.~\eqref{eq:appr:DPS:dampingE} invokes the damping parameter $\mathcal{S}[E^*]$ defined in Eq.~\eqref{eq:appr:DPS:dampingS}. The two parameters we have taken to define the damping coefficient are $E_0=20$ MeV and $E_1=15$ MeV. As previously mentioned, the excitation energy dependency of the temperature is taken at the Thomas-Fermi approximation to be
\begin{equation}
    T(\MQS{\chi}) = \left[\frac{E^*(\MQS{\chi})}{a}\right]^{1/2}\ ,
\end{equation}
with the nuclear level density parameter $a=A/8$.
To obtain an implementation equivalent to the state-of-the-art random-walk, we have set $\Delta U_{\rm thresh.}=0$ MeV. Recall that this threshold does not directly appear in the standard formalism of the random-walk and corresponds to the energy difference $\Delta U$ between two neighboring lattices below which a transition $\mathcal{I}_k\rightarrow\chi_k$ is certain in the Markov Chain~\figref{fig:appr:DPS:RW}. The potential energy surface is calculated on a regular grid following the work of Ref.~\cite{moller2009heavy}. Two points $\MQS{\chi}$ and $\MQS{\chi}'$ are neighbors if all the integer coordinates on the lattice differ by at most one unit. Such a definition in five dimensions leads to a maximum of $3^5-1=242$ neighbors for each node of the grid.

Our initial distribution is chosen to be entirely on the lattice site at the bottom of the PES' first potential well. The determination of this point is obtained as follow:
\begin{enumerate}
    \item we start at the origin of the lattice associated with the smallest elongation, a maximal neck radius, and spherical left and right bodies of the same volume;
    \item we iterate over the neighbors of the points, we select the neighbor that is associated with the lowest energy, and we reiterate up to reaching a local minimum $\MQS{\chi}_{\rm loc}$;
    \item we determine the minimum energy $E_{\rm sad.}$ required to reach scissioned configurations from $\MQS{\chi}_{\rm loc}$;
    \item we calculate the set $\mathcal{C}$ of all the configurations accessible from $\MQS{\chi}_{\rm loc}$ with an energy lower than $E_{\rm sad.}$, and define the bottom of the first potential well as the node $\MQS{\chi}_{\rm g.s.}\in\mathcal{C}$ associated with the lowest energy $E_{\rm g.s.}$. 
\end{enumerate}
By using this procedure, we also obtain the saddle energy $E_{\rm sad}$, as well as an effective barrier height $E_{\rm B} = E_{\rm sad} - E_{\rm g.s.}$. A site of the lattice is a scissioned configuration if its corresponding sharp macroscopic density has a neck radius $r_{\rm neck} < r_{\rm sciss.}$. We have calculated the fragment probability distribution for
\begin{equation}
    r_{\rm sciss} = 1.75,2.25,2.75\text{ fm} 
\end{equation}
and three different excitation energies $E^*$ such that
\begin{equation}
    x = E^* - E_{\rm B} = 0.1,2.0,4.0\text{ MeV,} 
\end{equation}
for a total of nine fission calculations per nucleus.

\subsection{Results}
\label{subsec:appli:results}

We have used our approach to calculate the fission fragment charge and mass probability distribution before prompt emission for the reactions \textsuperscript{233}U(n,f) and \textsuperscript{235}U(n,f).
\begin{figure}[b]
 \centering
  \includegraphics[width=0.99\linewidth]{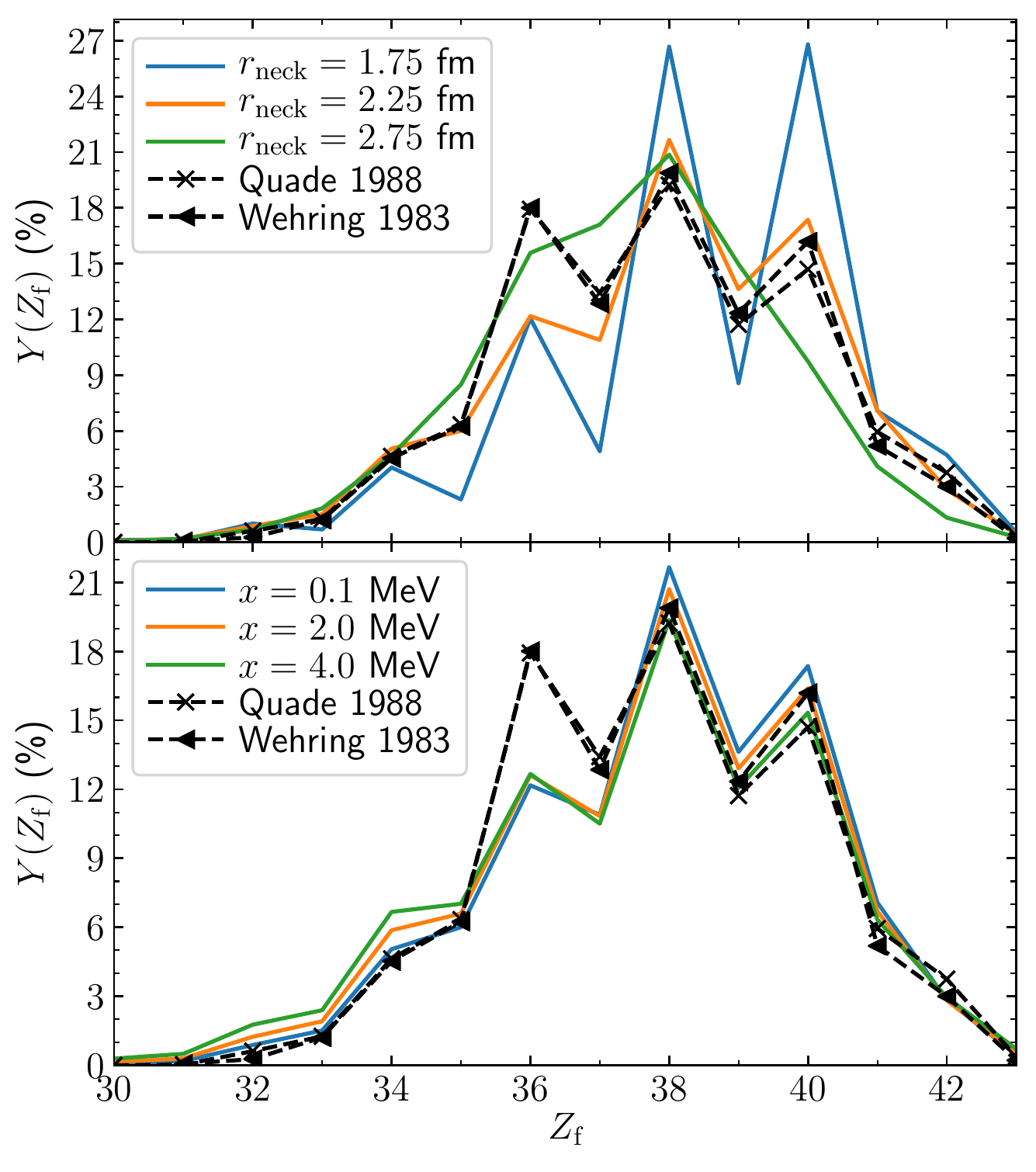}
  \caption{
    Fragment charge yields obtained with our approach for the reaction \textsuperscript{233}U($n_{\rm th.}$,f) at $x=0.1$ MeV with different neck radii (top panel) and with $r_{\rm neck}=2.25$ fm at different excitation energies (bottom panel). Our results are compared with experimental data with thermal incident neutron energies ($E_{\rm neut.}=0.0253$ eV) from~\cite{quade1988nuclide,wehring1980light}. \label{fig:appli:results:YZ-234U}
  }
\end{figure}
\begin{figure}[b]
 \centering
  \includegraphics[width=0.99\linewidth]{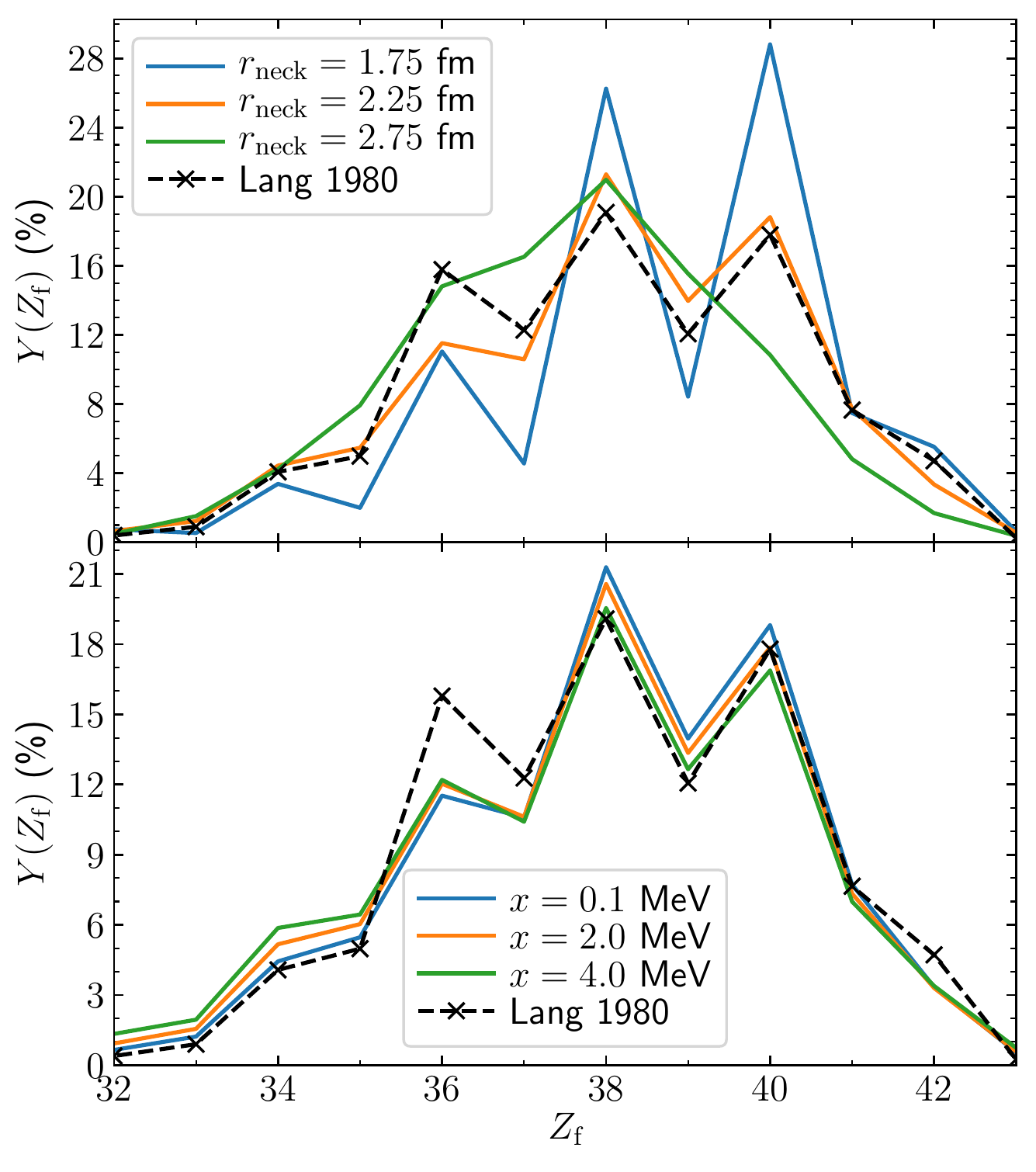}
  \caption{
    Fragment charge yields obtained with our approach for the reaction \textsuperscript{235}U($n_{\rm th.}$,f) at $x=0.1$ MeV with different neck conditions (top panel) and with $r_{\rm neck}=2.25$ fm at different excitation energies (bottom panel). Our results are compared with experimental data with thermal incident neutron energies ($E_{\rm neut.}=0.0253$ eV) from~\cite{quade1988nuclide,wehring1980light}. \label{fig:appli:results:YZ-236U}
  }
\end{figure}
We highlight in Figs. \ref{fig:appli:results:YZ-234U} and \ref{fig:appli:results:YZ-236U} the charge yields of these reactions respectively.
The most interesting aspect of these calculations is the appearance of an odd-even staggering in the fragment's charge distributions. 
The odd-even staggering is commonly attributed to pairing effects, since separating paired particles is energetically costly, but also from structural effects (see, e.g., Refs~\cite{caamano2011evidence,gonnenwein2013evenodd,moller2014fission,jurado2015influence} and references within). 

We see a strengthening of the odd-even staggering for the smallest neck radius considered in our calculations. 
To explain this phenomenon, already observed in Ref.~\cite{verriere2019number}, consider a system with two clearly separated fragments where the distance is great enough so that the interaction between the fragments is minuscule. 
Neglecting pairing for the sake of simplicity, the mean-field Hamiltonian can be diagonalized in a particle basis of states that are localized in only one of the fragments. 
Since we minimize the total binding energy, each fragment must have a well-defined number of particles. 
Preservation of time-reversal symmetry ensures each particle state is associated with a time-reversal state with the same energy, the same spatial distribution, and the same occupation (Kramers degeneracy theorem). 
Since each fragment must have a well-defined number of particles and each particle must have a time-reversal partner, each fragment can only possess an even number of particles, thus reducing the odd-proton abundance. 
In conclusion, if (i) we assume the mean-field approximation, (ii) we minimize the total binding energy of the system, and (iii) we preserve the time-reversal symmetry (as in our approach), then two separated fragments can only have an even number of particles, leading to a strengthening of the odd-even staggering of the fragments when the separation distance is too large (conversely, neck radius too small). 

The lower panels of~\figref{fig:appli:results:YZ-234U} and~\figref{fig:appli:results:YZ-236U} exhibit very little variation in our charge yields with excitation energy $0.1-4.0$ MeV above the fission barrier. 
Data in this energy range confirm a small variation of the yields at low excitation energy \cite{gooden2016energy, shibata2011jendl}. 
In this work, we determine our fragmentation probabilities for each scissioned configuration, assuming zero-temperature microscopic states. 
Therefore, the energy dependence in our current predictions only comes from the potential energy evolution along the collective variables and discrepancies could arise at higher incoming neutron energies. 
Despite this approximation, the reactions studied in this work exhibit an excellent quantitative agreement with thermal-neutron experimental data for nearly all proton numbers. 
The agreement is particularly good when the scission criteria is set to $r_{\rm neck}=2.25$ fm. 
We retain this value for the remainder of our results. 

\begin{figure}[tb]
 \centering
  \includegraphics[width=0.99\linewidth]{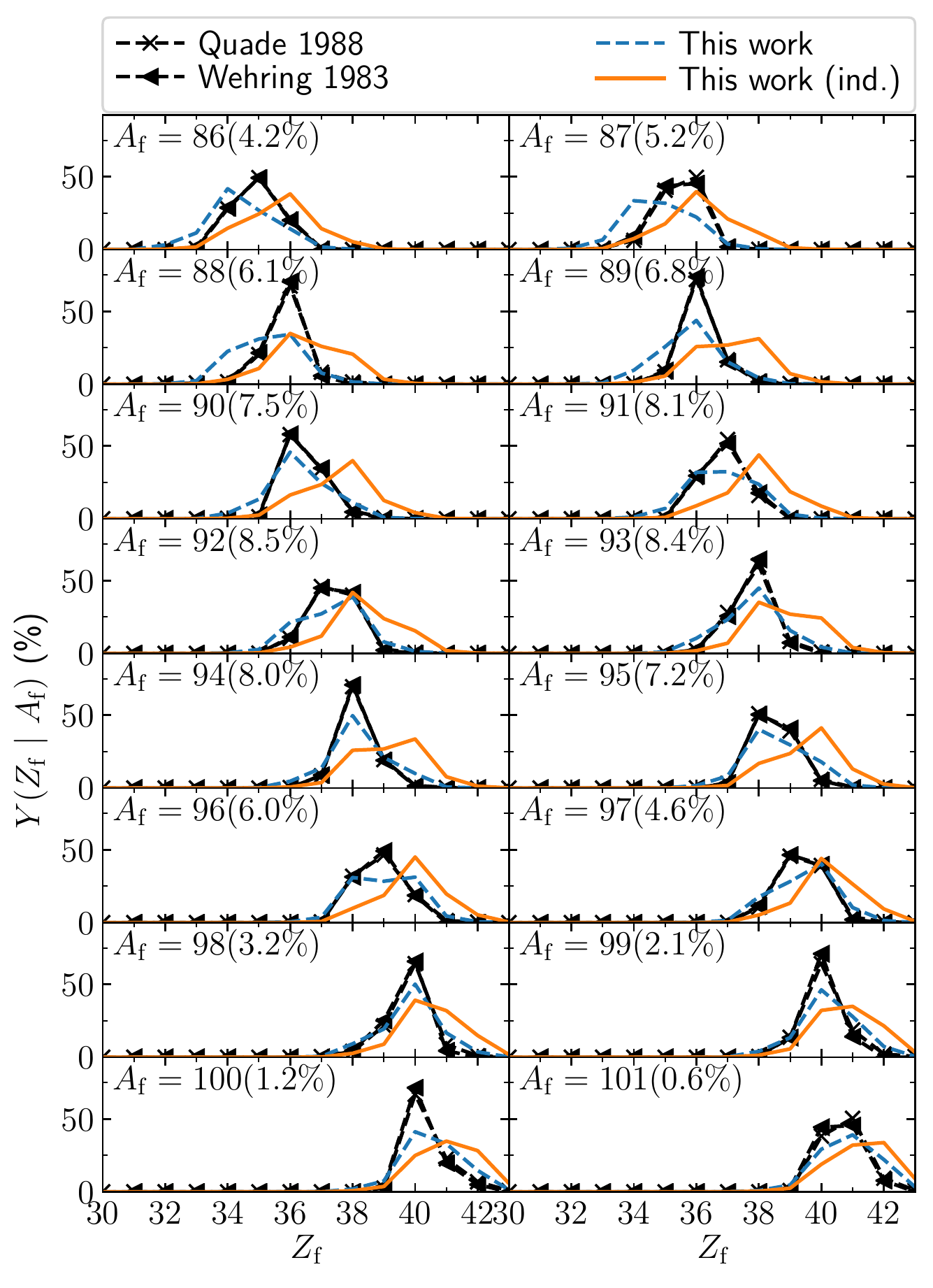}
  \caption{Isobaric fragment charge yields (dashed blue lines) and the independent charge yields (solid orange lines), after prompt neutron emission, obtained with our approach for the reaction \textsuperscript{233}U($n_{\rm th.}$,f) with a neck condition $z_{\rm neck}=2.25$ fm and an excitation energy above the barrier of $x=0.1$ fm. Each curve is normalized at the fixed $A$ to 100\% and the fragment probability associated with each of the masses is listed in parentheses. We compare our results with experimental independent yields (dashed black curve) from Refs.~\cite{quade1988nuclide, wehring1980light}.
  \label{fig:appli:results:YAZ-234U}
  }
\end{figure}

\begin{figure}[tb]
 \centering
  \includegraphics[width=0.99\linewidth]{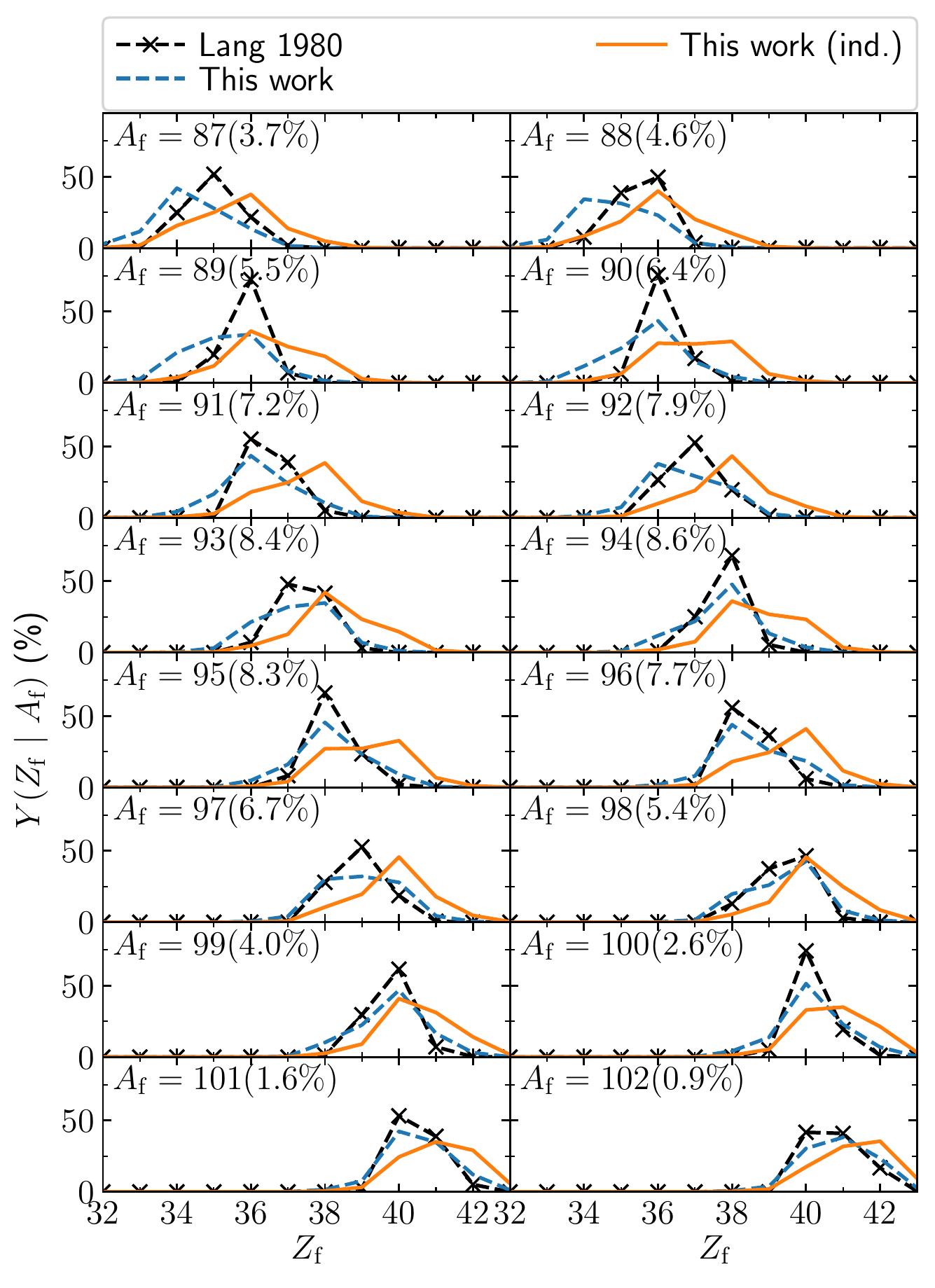}
  \caption{Same as \figref{fig:appli:results:YAZ-234U}, where the experimental independent yields are this time taken from Ref.~\cite{lang1980nuclear}.
  \label{fig:appli:results:YAZ-236U}
  }
\end{figure}

As previously mentioned, our approach enables the determination of the full fragment yields $Y(\Zfrag,\Afrag)$. 
We first present our charge yields $Y(\Zfrag\,|\,\Afrag)$ for different fixed fragment masses $\Afrag$
\begin{equation}
    Y(\Zfrag\,|\,\Afrag) = \frac{Y(\Zfrag,\Afrag)}{Y(\Afrag)}
\end{equation}
for the reaction \textsuperscript{233}U(n,f) in~\figref{fig:appli:results:YAZ-234U} and for the reaction \textsuperscript{235}U(n,f) in~\figref{fig:appli:results:YAZ-236U}. 
We approximate independent yields (solid orange lines) from our calculated fragment yields operating under the assumption that the prompt neutron emission multiplicity distribution, $P_\nu(\Afrag)$, depends on the mass of the fragments only. 
To estimate $P_\nu(\Afrag)$, we first fit the parameter $p$ of the probability density function of the binomial random variable $\mathcal{B}(N=5,p)$ on the experimental distribution to emit $\nu$ neutrons from any fragment from Ref.~\cite{gwin1984measurements} as suggested in~\cite{diven1956multiplicities}. 
We then shift this distribution for each fragment mass such that the expected value of emitted neutrons is equal to $\bar{\nu}(\Afrag)$ from Refs.~\cite{nishio1998multi, nishio1998thermal} to obtain $P_\nu(\Afrag)$. 
This method is easy to implement and fast, albeit simple; providing a means for comparison with experimental independent yield data. 
The presented theoretical yields correspond to an excitation energy $x=0.1$ MeV above the barrier. 
As previously mentioned, our calculations show a slow evolution of the yield as a function of excitation energy so the other results at higher excitation energy will be qualitatively similar. 
In these two figures, the fragment isobaric yields are closer to experimental data than the independent ones (generally shifted toward a higher charge number than experimental data). 
Possible explanations for this discrepancy include the assumption of strongly damped motion, or a lack of optimization in our mac-mic model parameters. 
If the latter of these hypotheses is true, it reinforces the need to optimize model parameters in conjunction with improvements to the physics of the model. 

\begin{figure}[ht]
 \centering
  \includegraphics[width=0.99\linewidth]{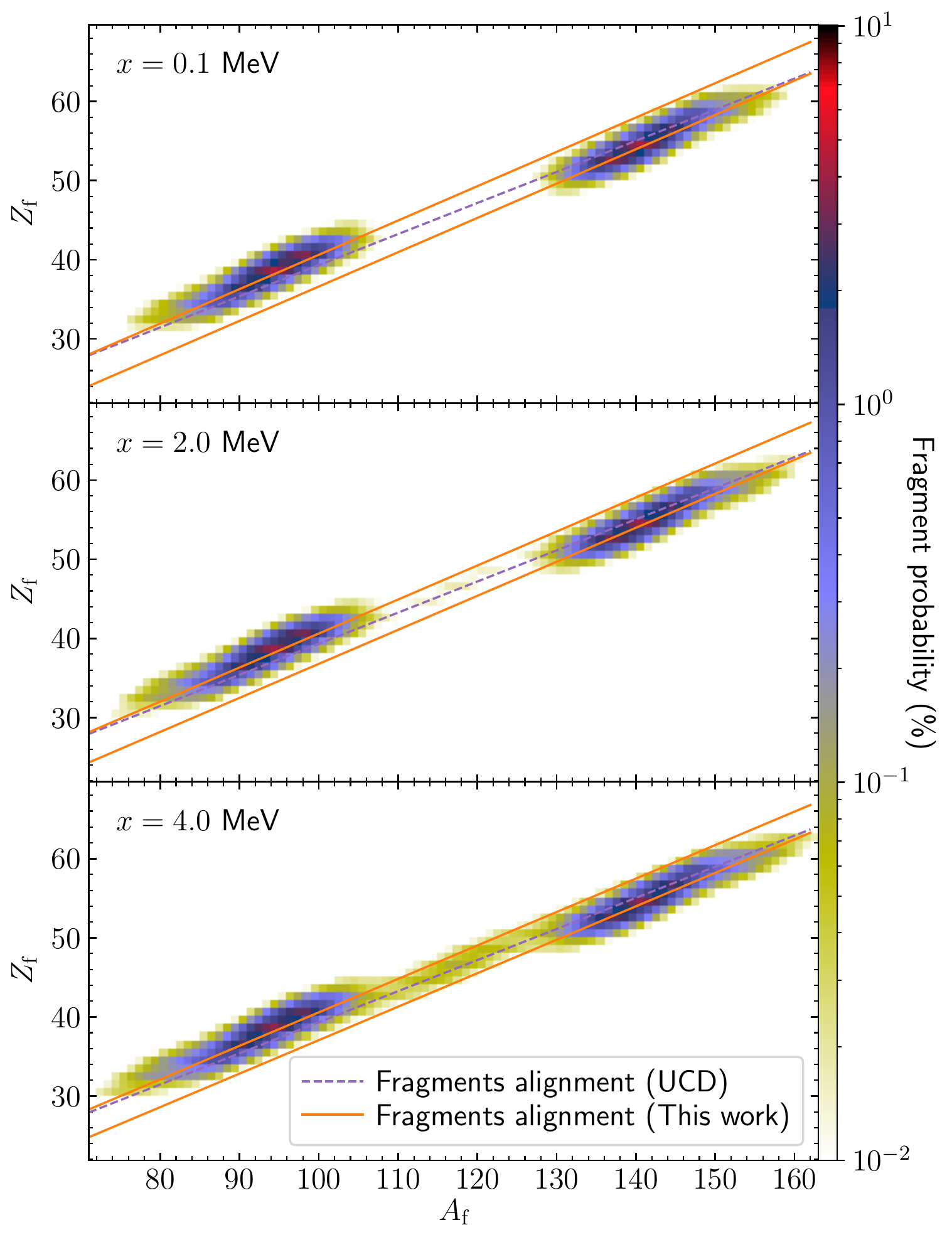}
  \caption{Fragment mass and charge yields, before prompt neutron emission, obtained with our approach for the reaction \textsuperscript{233}U($n_{\rm th.}$,f) with a neck condition $z_{\rm neck}=2.25$ fm. Each panel shows the yield with different excitation energies above the barrier. The alignment of the fragments (blue lines) is obtained through a linear fit of $\bar{\Zfrag}(\Afrag)$. The green line corresponds to the alignment the fragments under the UCD approximation.
  \label{fig:appli:results:YAZ-imshow-234U}
  }
\end{figure}

\begin{figure}[ht]
 \centering
  \includegraphics[width=0.99\linewidth]{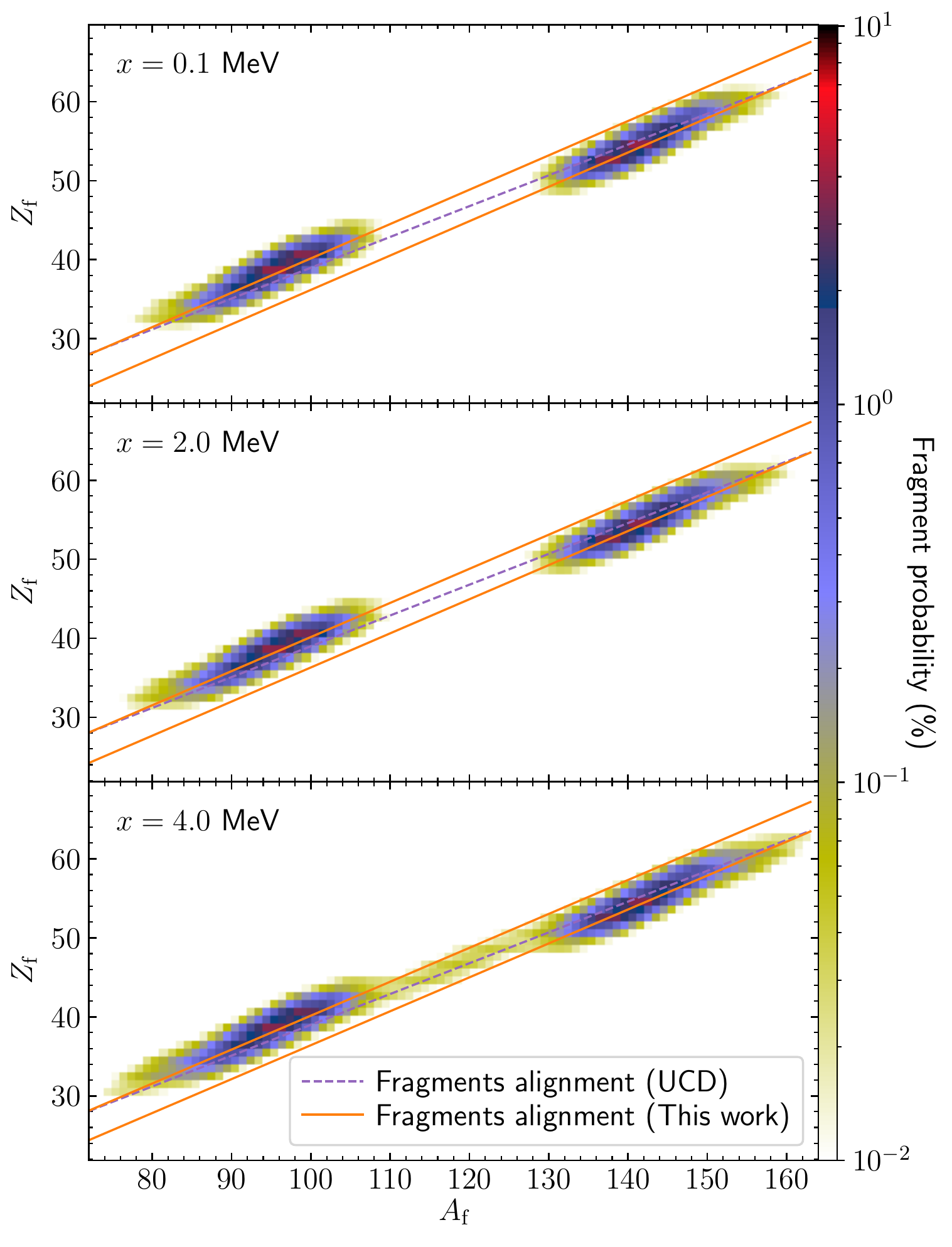}
  \caption{Fragment mass and charge yields, before prompt neutron emission, obtained with our approach for the reaction \textsuperscript{235}U($n_{\rm th.}$,f) with a neck condition $z_{\rm neck}=2.25$ fm. Each panel shows the yield with different excitation energies above the barrier. The alignment of the fragments (blue lines) is obtained through a linear fit of $\bar{\Zfrag}(\Afrag)$. The green line corresponds to the alignment the fragments under the UCD approximation.
  \label{fig:appli:results:YAZ-imshow-236U}
  }
\end{figure}

\begin{figure}[tb]
 \centering
  \includegraphics[width=0.99\linewidth]{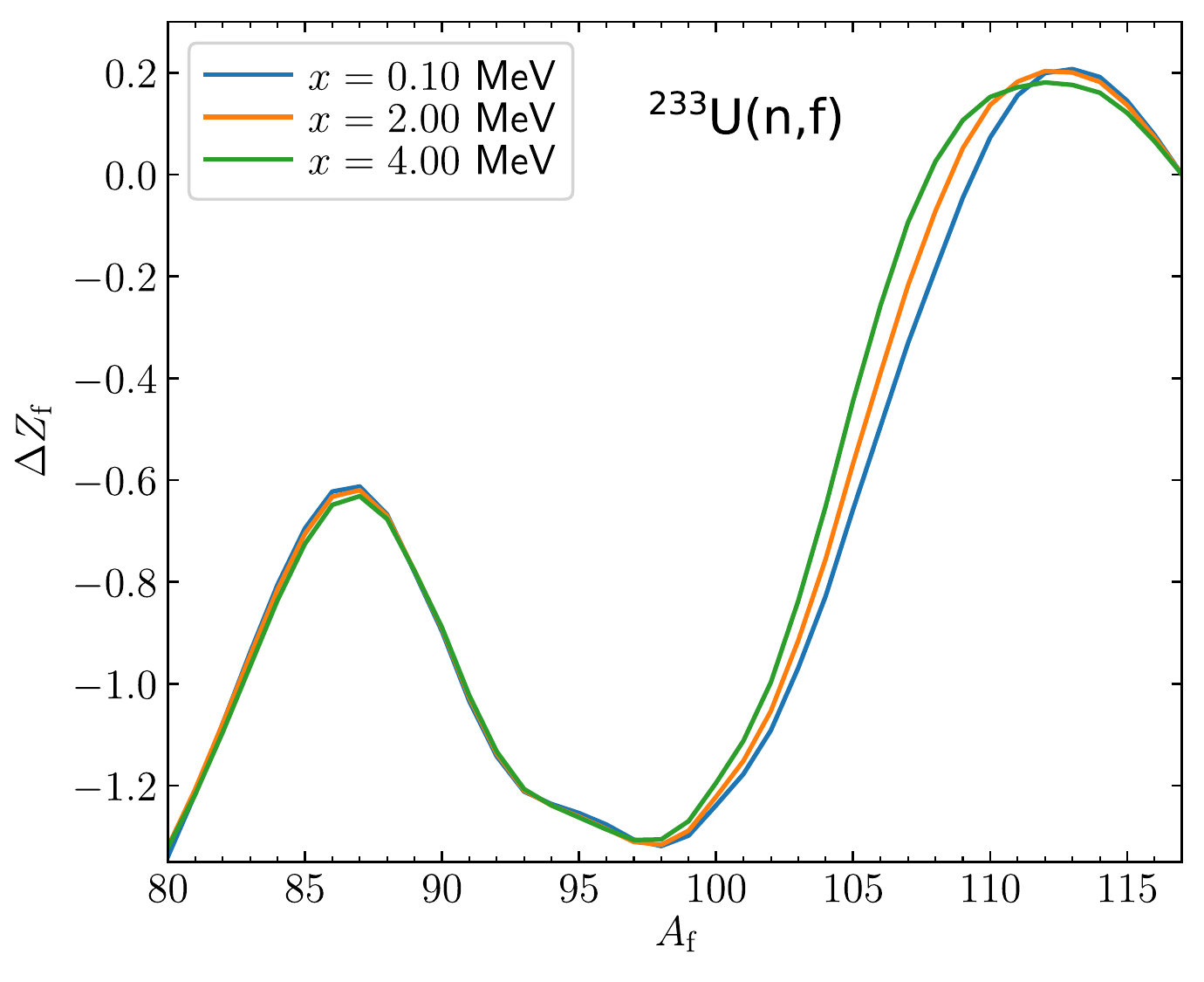}
  \includegraphics[width=0.99\linewidth]{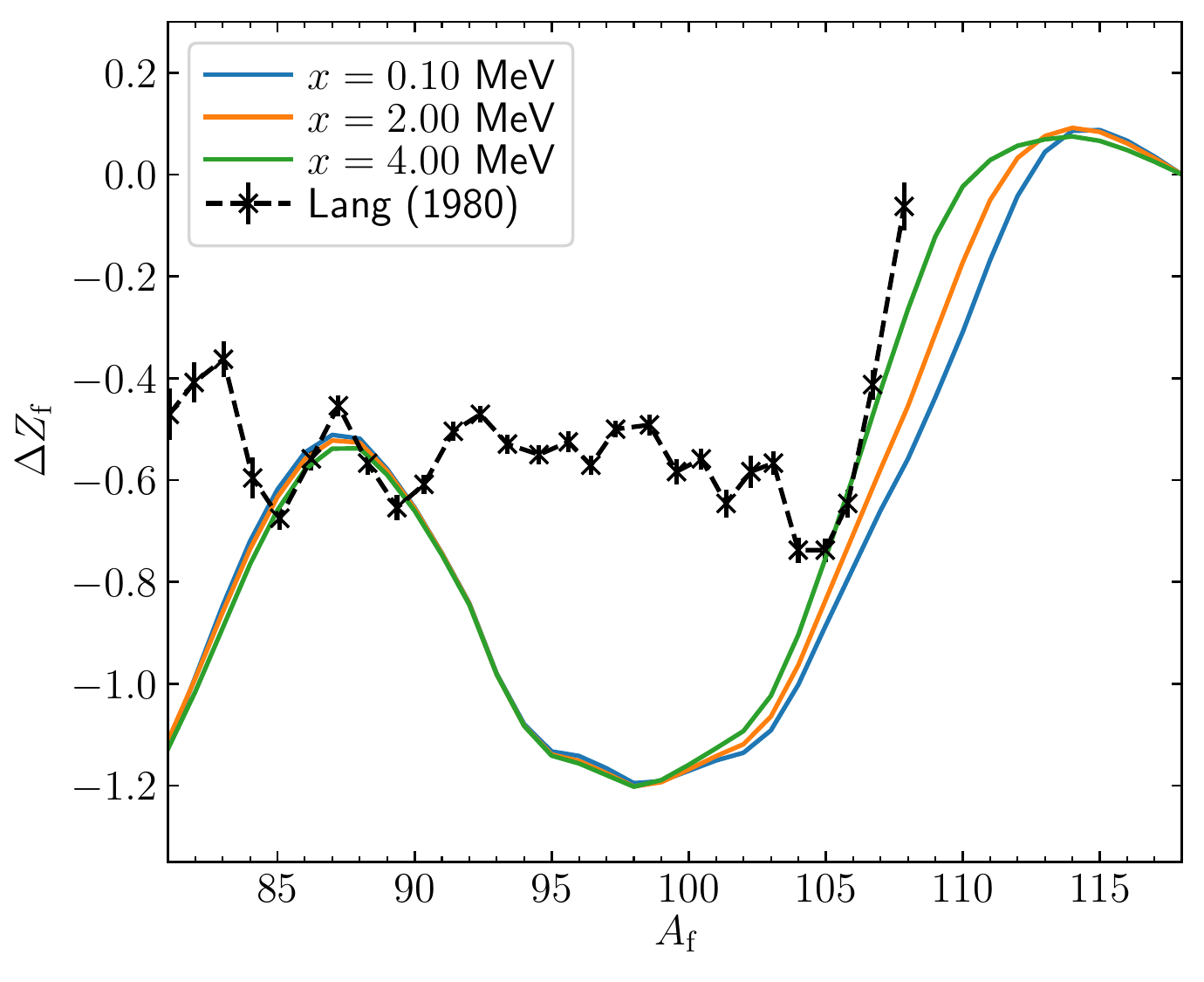}
  \caption{Deviation of $\bar{Z}$ from the Unchanged Charge Distribution (UCD) approximation in our calculation, for each fragment charge $Z$ and different excitation energies above the barrier $E_x$. Our results for the reaction $^{235}$U(n, f) is compared with experimental data from Ref.~\cite{lang1980nuclear}. We see a substantial deviation from the UCD approximation.
  \label{fig:appli:results:mpc-dev}
  }
\end{figure}

\begin{figure}[tb]
 \centering
  \includegraphics[width=0.99\linewidth]{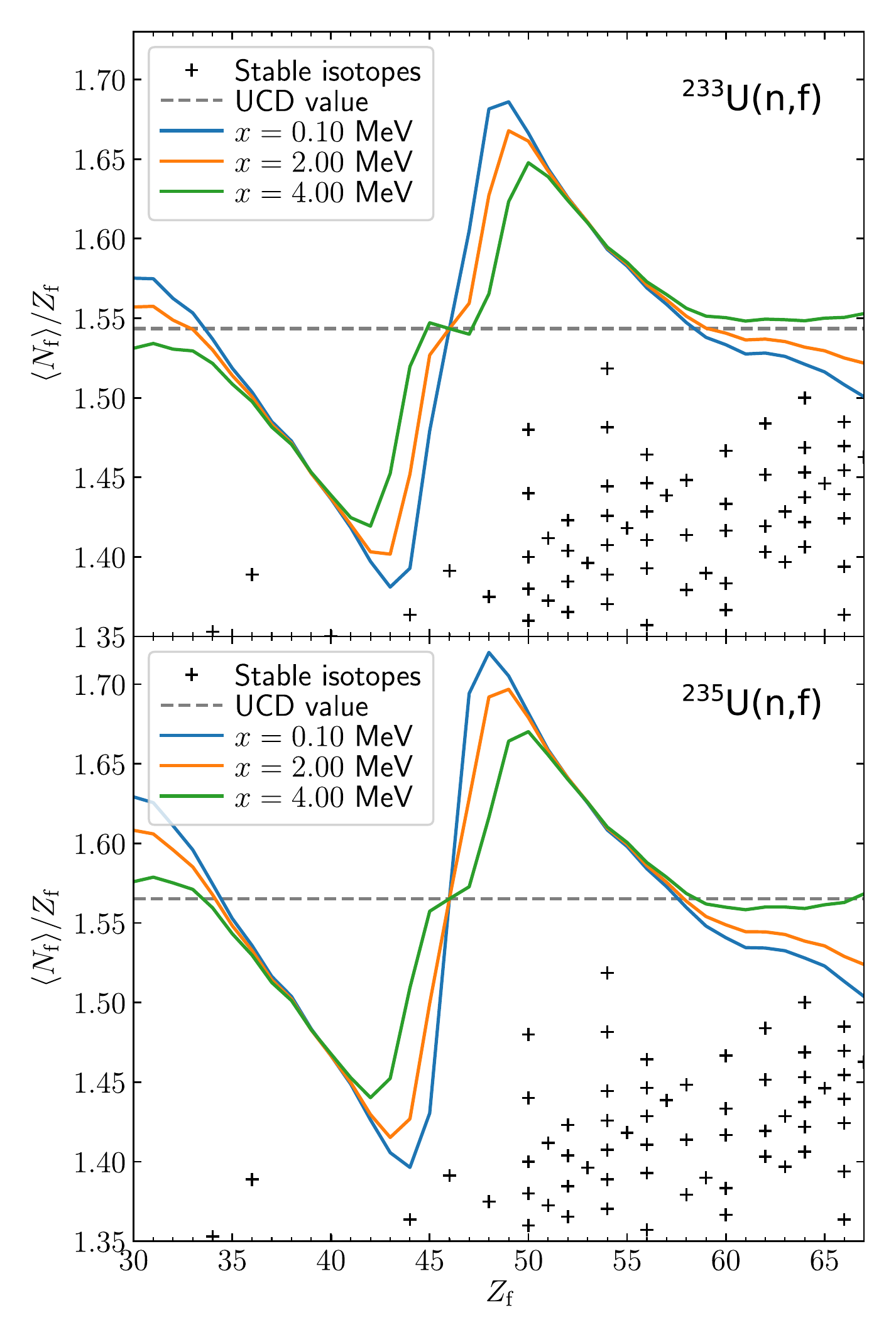}
  \caption{Neutron excess in our calculation, for each fragment charge $Z$ and different excitation energies above the barrier $E_x$. The dashed gray line corresponds to the value under the Unchanged Charge Distribution (UCD) approximation.
  \label{fig:appli:results:neut-excess}
  }
\end{figure}

Figures \ref{fig:appli:results:YAZ-imshow-234U} and \ref{fig:appli:results:YAZ-imshow-236U} present the full fragment mass and charge distribution we have obtained using our approach for the respective reactions \textsuperscript{233}U($n_{\rm th.}$,f) and \textsuperscript{235}U($n_{\rm th.}$,f). 
The main feature in these results is the emergence of a charge polarization of the fission fragments. 
Our results show a deviation of the mean primary charge
\begin{equation}\label{eq:results:dZ}
    \Delta \Zfrag(\Afrag) = \Zfrag^{\rm UCD}-\bar{\Zfrag}(\Afrag) \ ,
\end{equation}
where $\Zfrag(\Afrag)$ is the average number of charge obtained in our approach and $\Zfrag^{\rm UCD}(\Afrag)$ is the UCD value. 
Both quantities are given by
\begin{align}
    \bar{\Zfrag}(\Afrag) &= \frac{\sum_{\Zfrag} \Zfrag \times Y(\Zfrag, \Afrag)}{\sum_{\Zfrag} Y(\Zfrag, \Afrag)} \\
    \Zfrag^{\rm UCD}(\Afrag) &= \eta\Afrag = \frac{Z}{A}\Afrag\ .
\end{align}
Experimentally, $\Delta \Zfrag(\Afrag)$ has been measured to be $\Delta\Zfrag^{\rm exp.}\approx -0.6 \pm 0.2$ for the light fragments having $\Afrag=80,\dots,105$ in the case of the reaction $^{235}$U(n,f) at thermal neutron energies~\cite{lang1980nuclear, schmitt1984fission}. 
Figure \ref{fig:appli:results:mpc-dev} shows the deviation $\Delta \Zfrag(\Afrag)$ we have obtained with our approach. 
It shows a value of $\Delta \Zfrag(\Afrag)$ between $-0.5$ and $-1.2$ for the light fragments. 
The average neutron excess has also been experimentally obtained for other fission reactions, in inverse kinematic, in Refs.~\cite{caamano2015characterization,ramos2020scission}. 
Figure \ref{fig:appli:results:neut-excess} shows the neutron excesses we have obtained using our method for both reactions $^{233,235}$U(n,f). 
Qualitatively, we reproduce the overall shape of the neutron excess curves of experimental data obtained in other fissioning systems~\cite{caamano2015characterization, ramos2020scission}. 
For both reactions, the deviation from the UCD decreases with increasing excitation energy. 
The difference between the maximum and the minimum neutron excess in our results is around 0.27 at $E_x=2$ MeV and 0.22 at $E_x=4$ MeV while experimentally, this value drops to 0.12 for $^{240}$Pu and 0.09 for $^{239}$U. 

Several explanations of the physical origin of the charge polarization of the fission yields before prompt neutron emission have been proposed in the literature. 
For example, in Ref.~\cite{norenberg1966theory}, the authors propose three possible origins: (i) from the level density of the fragments, (ii) from the Coulomb interaction between the fragments and (iii) from the asymmetry energy. 
No experimental data were available at the time regarding the charge polarization of the fragment yields. 
In Ref.~\cite{caamano2015characterization}, the authors compare their experimental results with a scission-point Liquid drop model (LD-SP) able to reproduce the smooth trend of the experimental charge polarization, which leads to a possible decomposition of the charge polarization of the primary yields into a smooth component, whose origin lies in the asymmetry energy, as well as a sharp part, coming from microscopic shell and pairing effects. 

\subsection{Validity of the convergence criteria}
\label{subsec:appli:cvgval}

\begin{figure}[ht]
 \centering
  \includegraphics[width=0.99\linewidth]{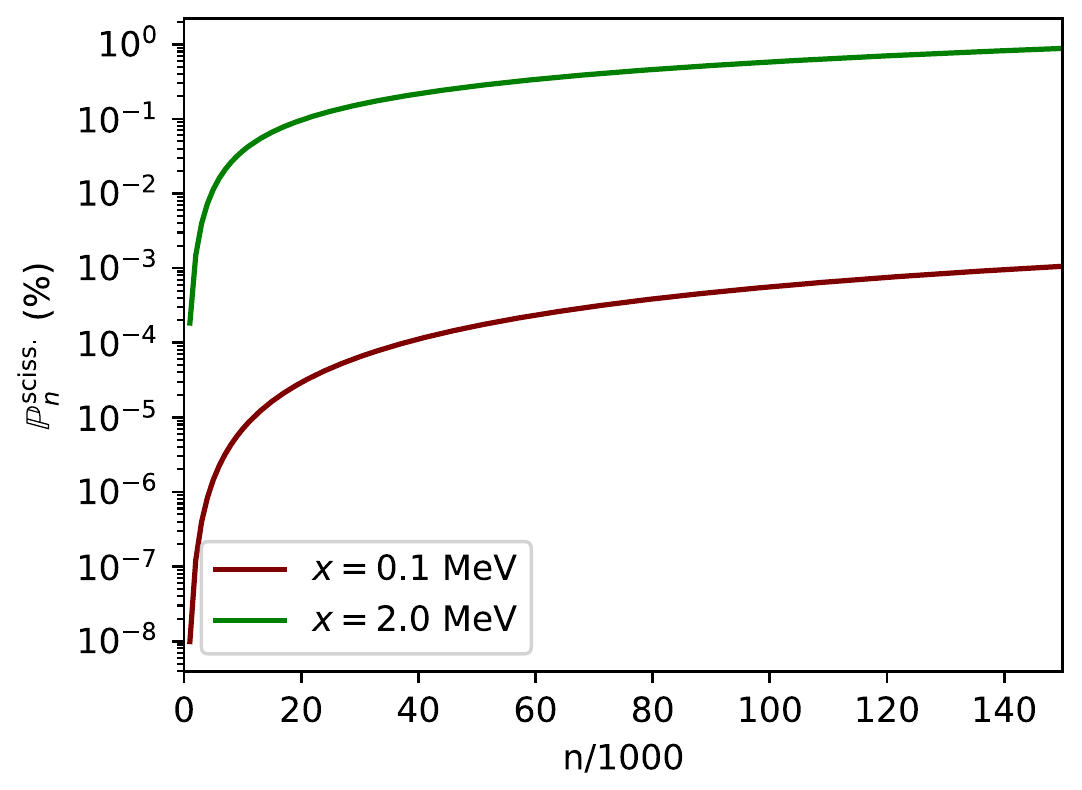}
  \caption{
    Evolution of the scission probability according to the number of MCMC random walk steps, determined with our novel algorithm, DPS, for the reaction \textsuperscript{233}U(n,f).\label{fig:appli:cvgval:Psciss}
  }
\end{figure}

\begin{figure}[ht]
 \centering
  \includegraphics[width=0.99\linewidth]{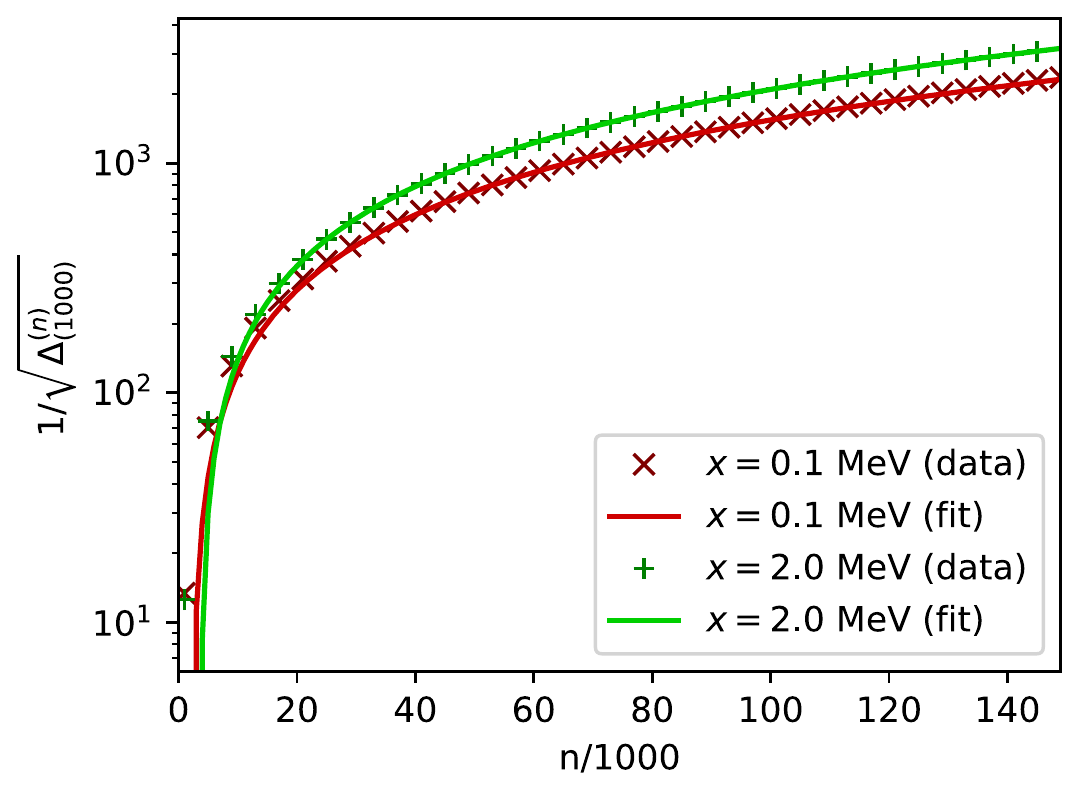}
  \caption{
    Inverse-square-root of the evolution of the distance between successive probability distributions of the scissioned configurations for the reaction \textsuperscript{233}U(n,f). To keep a readable graph, we have displayed only one point over four. A linear fit of the data is presented for comparison purposes.\label{fig:appli:cvgval:isrde}
  } 
\end{figure}

To analyze the convergence properties of our new algorithm, we calculate the probability distributions of the scissioned configurations $a^{(n)}(\MQS{\chi})$ for all steps, $n \leq n_{\rm max} = 150,000$, for the reaction \textsuperscript{233}U(n,f). From this calculations, we can extract the probability to reach scission after $n$ iterations
\begin{equation}
  \mathbb{P}^{\rm sciss.}_n =
    \sum_{\MQS{\chi}}\bar{a}^{(n)}(\MQS{\chi}) \ .
\end{equation}
The evolution of this quantity with increasing $n$ is shown in~\figref{fig:appli:cvgval:Psciss}. As expected, the lower excitation energy (red curve) is significantly below the run with higher excitation energy (green curve). 
For the same range of iterations, the evolution of the inverse-square-root of $\Delta_{(1000)}^{(n)}$ is presented in~\figref{fig:appli:cvgval:isrde}. Recall that $\Delta_{j}^{(n)}$ is the convergence error given by Eq.~\eqref{eq:appr:cc}.
The inverse-quadratic regime for $\Delta_{(1000)}^{(n)}$ is reached around $n\approx 20,000$ iterations where the ratio of paths that are reaching scission is $3.0\times 10^{-5}\%$ for $x=0.1$ MeV and $0.096\%$ for $x=2.0$ MeV. In fact, the convergence seems to be at least quadratic, meaning that our convergence criteria is overestimating the error. 

\begin{figure}[ht]
 \centering
  \includegraphics[width=0.99\linewidth]{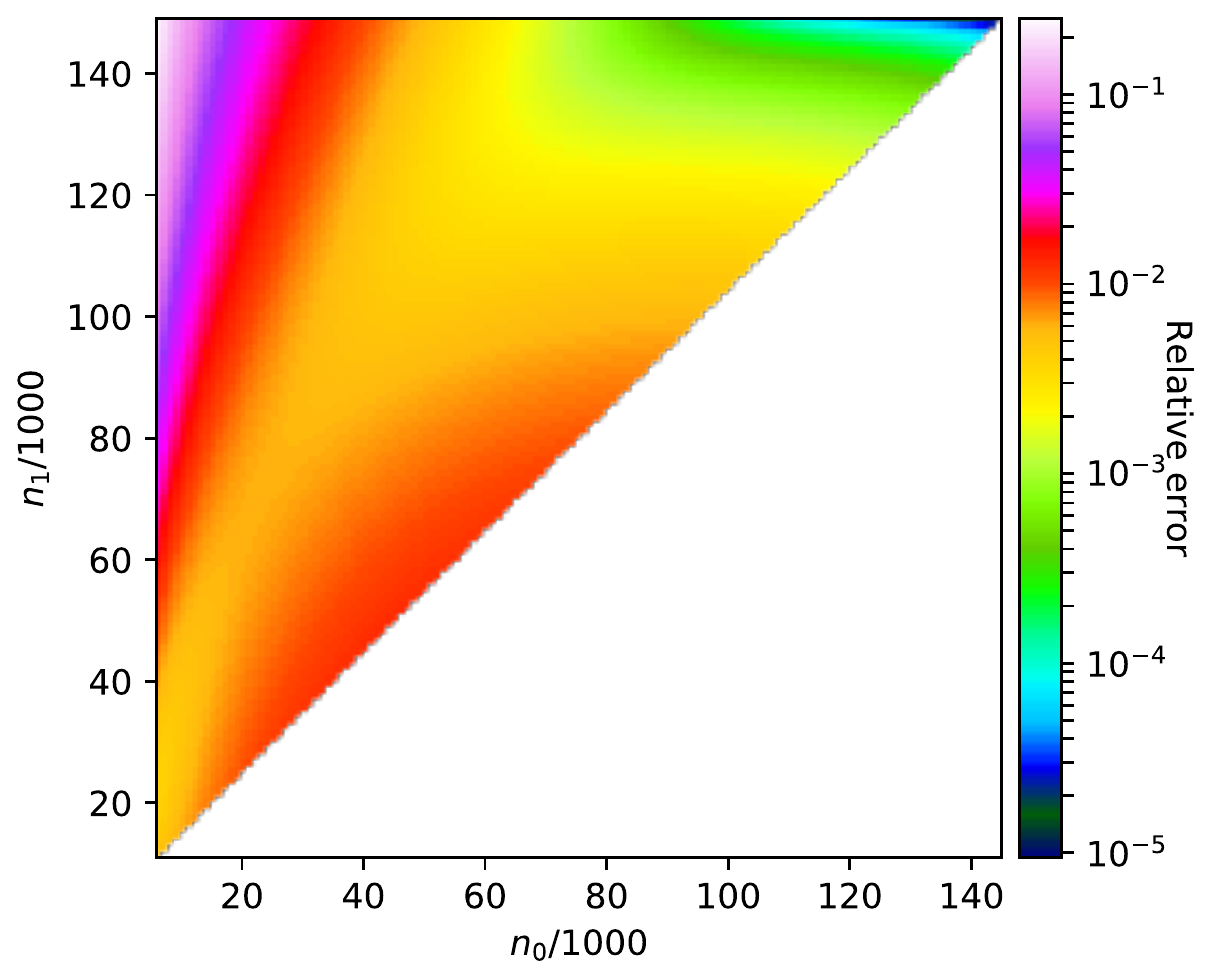}
  \includegraphics[width=0.99\linewidth]{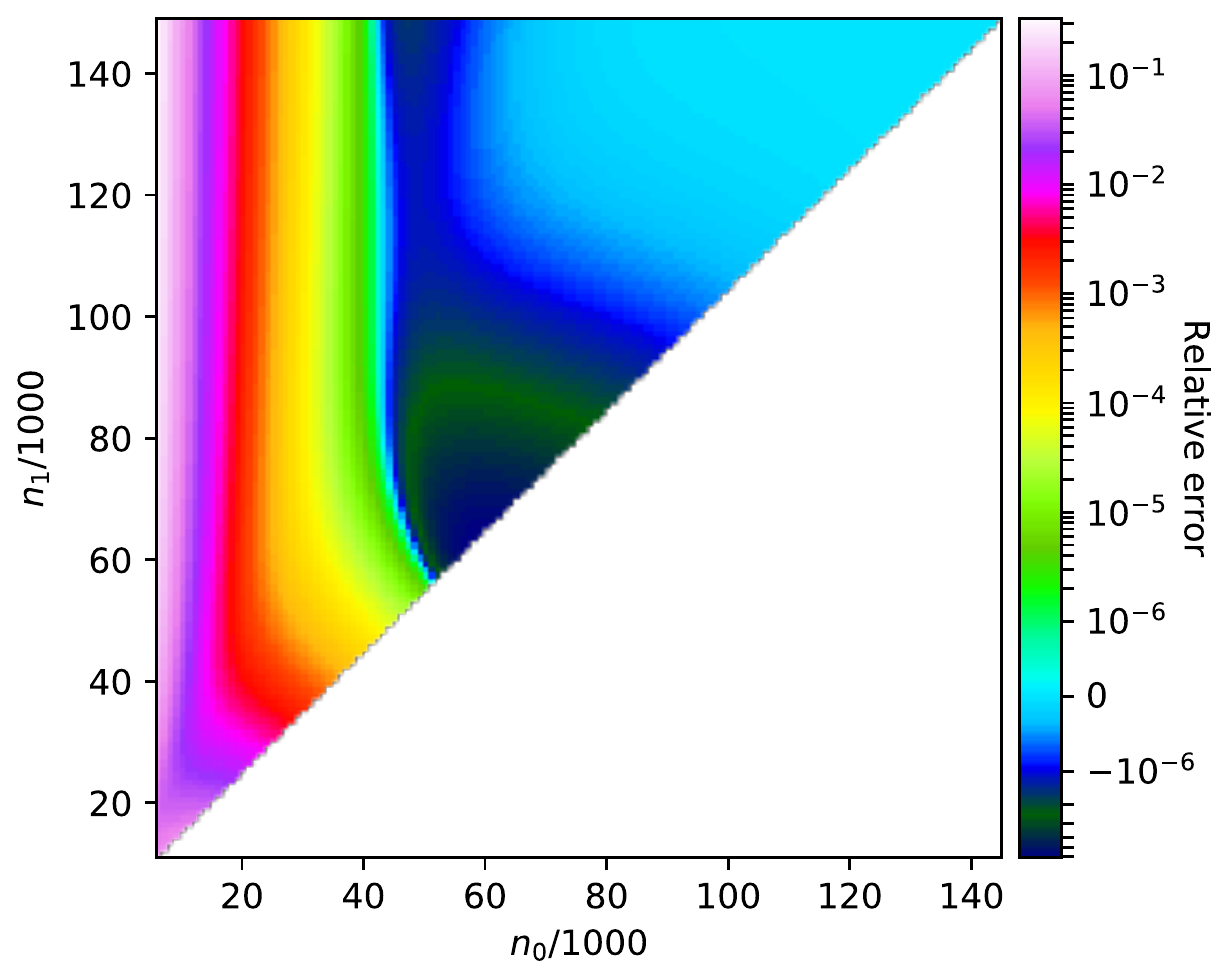}
  \caption{
    Relative error $D_{n_0,n_1}$, defined by Eq.~\eqref{eq:appli:cvgval:relerr}, between our convergence criteria and the exact convergence error, for each interval $n_0,\dots,n_1$ of fitted data, for the reaction \textsuperscript{233}U(n,f) at an excitation energy above the barrier of $x=0.1$ MeV (top panel) and $x=2.0$ MeV (bottom panel). \label{fig:appli:cvgval:cmp2D}
  }
\end{figure}

To further validate our convergence criteria, $\varepsilon^{(n,1000)}_{\rm eff.}$, we have estimated it through a fit on the data applicable in the range of iterations $n_0,\dots,n_1$. We compare our convergence criteria against data through the calculation of the relative error of the convergence criteria using $j=1000$
\begin{equation}\label{eq:appli:cvgval:relerr}
  D_{n_0,n_1} = 
    \frac{\displaystyle
        \left[
            \left.\varepsilon^{(n_0          ,j)}_{\rm eff.}\right|_{n_0}^{n_1}
          - \left.\varepsilon^{(n_{\rm max}+1,j)}_{\rm eff.}\right|_{n_0}^{n_1}
        \right]
      - \!\!\!\!
        \sum_{k=0}^{\frac{n_{\rm max}-n_0}{j}}
        \!\!\!
          \Delta_{(j)}^{(n_0+jk)}
    }{
        \left.\varepsilon^{(n_0          ,j)}_{\rm eff.}\right|_{n_0}^{n_1}
      - \left.\varepsilon^{(n_{\rm max}+1,j)}_{\rm eff.}\right|_{n_0}^{n_1}
    },
\end{equation}
where $\left.\varepsilon^{(n,j)}_{\rm eff.}\right|_{n_0}^{n_1}$ corresponds to the estimation of our convergence criteria using a fit between $n_0$ and $n_1$.
The evolution of $D_{n_0,n_1}$ according to the range used for the fit is presented~\figref{fig:appli:cvgval:cmp2D} for which $n_0>5,000$ and $n_1-n_0\geq 5,000$.
The relative error $D_{n_0,n_1}$ associated with our convergence criteria between the iterations $n_0$ and $n_1$ is below two percent after only ten thousand iterations in both cases. $D_{n_0,n_1}$ is greater than zero almost everywhere, which means that our criteria slightly overestimates the convergence error.

\subsection{Comparison with the Metropolis implementation}
\label{subsec:appli:DPSvsRW}

We compare our new calculations to past work in order to show that we can reproduce these efforts within the context of our more general methodology. 
We use the implementation of a discrete random walk (DRW) as in Ref.~\cite{mumpower2020primary} as the baseline FRLDM mass yield calculations.  
We perform these calculations for the reaction \textsuperscript{233}U(n,f) at an excitation energy of $x \approx 4$ MeV above the barrier; noting that this value takes into account differences between the potential energy surface used in Ref.~\cite{mumpower2020primary} and the potential energy surface of this work. 
The starting point of the DRW calculation is the ground state as chosen by the procedure of Ref.~\cite{mumpower2020primary}. 
The biased potential is set to zero in the DRW calculation and we set the scission neck radius to be $2.25$ fm. 
We accumulate 100,000 scissioned configurations for this comparison. 
With these parameters, we have nearly identical inputs as our new results shown in Sec.~\ref{subsec:appli:results}. 

\begin{figure}[ht]
 \centering
  \includegraphics[width=0.99\linewidth]{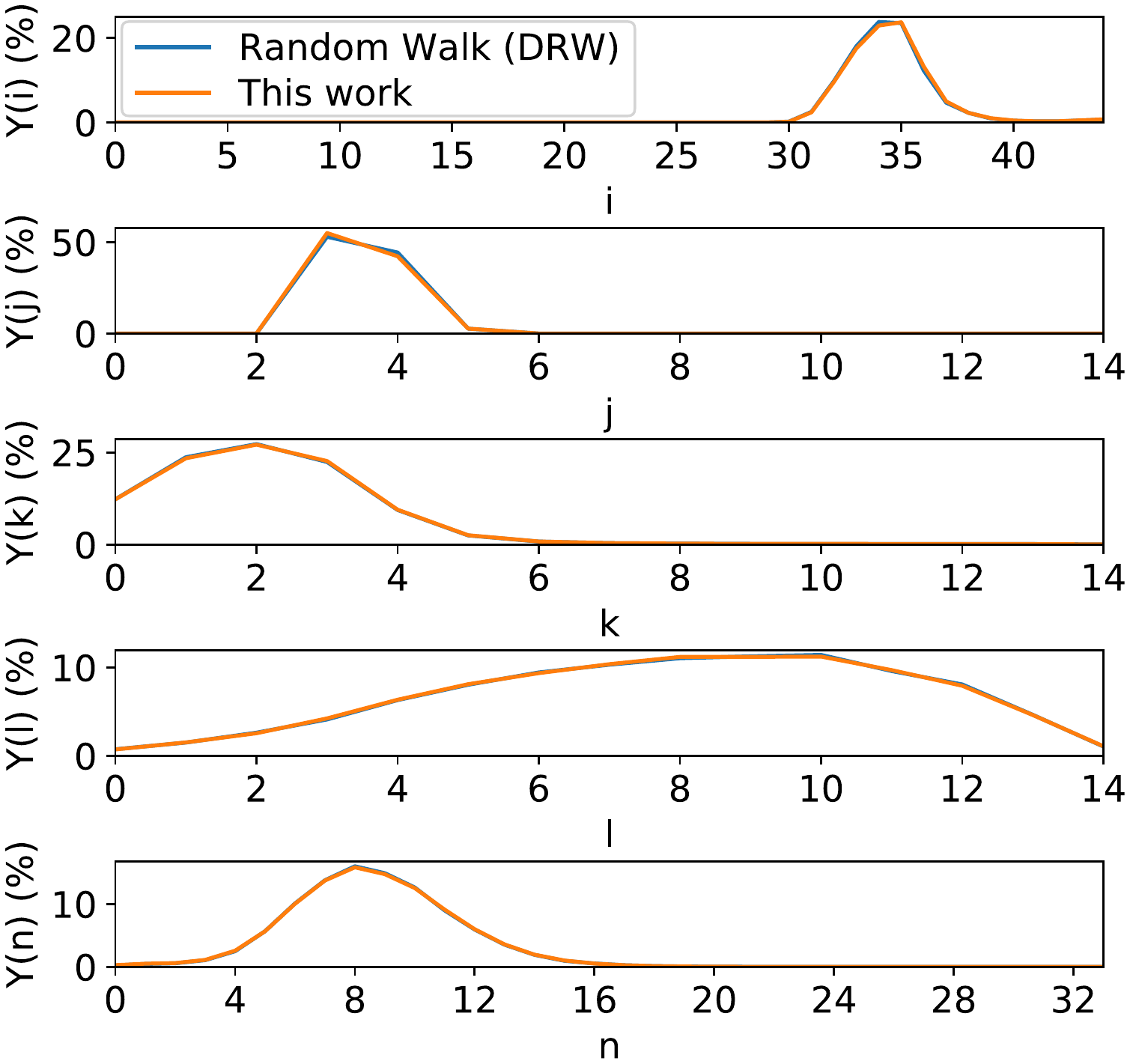}
  \caption{
    Probability distributions to obtain a scissioned configuration associated with each integer indices $(i,j,k,l,n)$ as defined in~\cite{moller2009heavy} estimated with state-of-the-art random walk with the code DRW (blue) and with this work (orange).\label{fig:appli:DPSvsRW:cmp}
  }
\end{figure}

\begin{figure}[ht]
 \centering
  \includegraphics[width=0.99\linewidth]{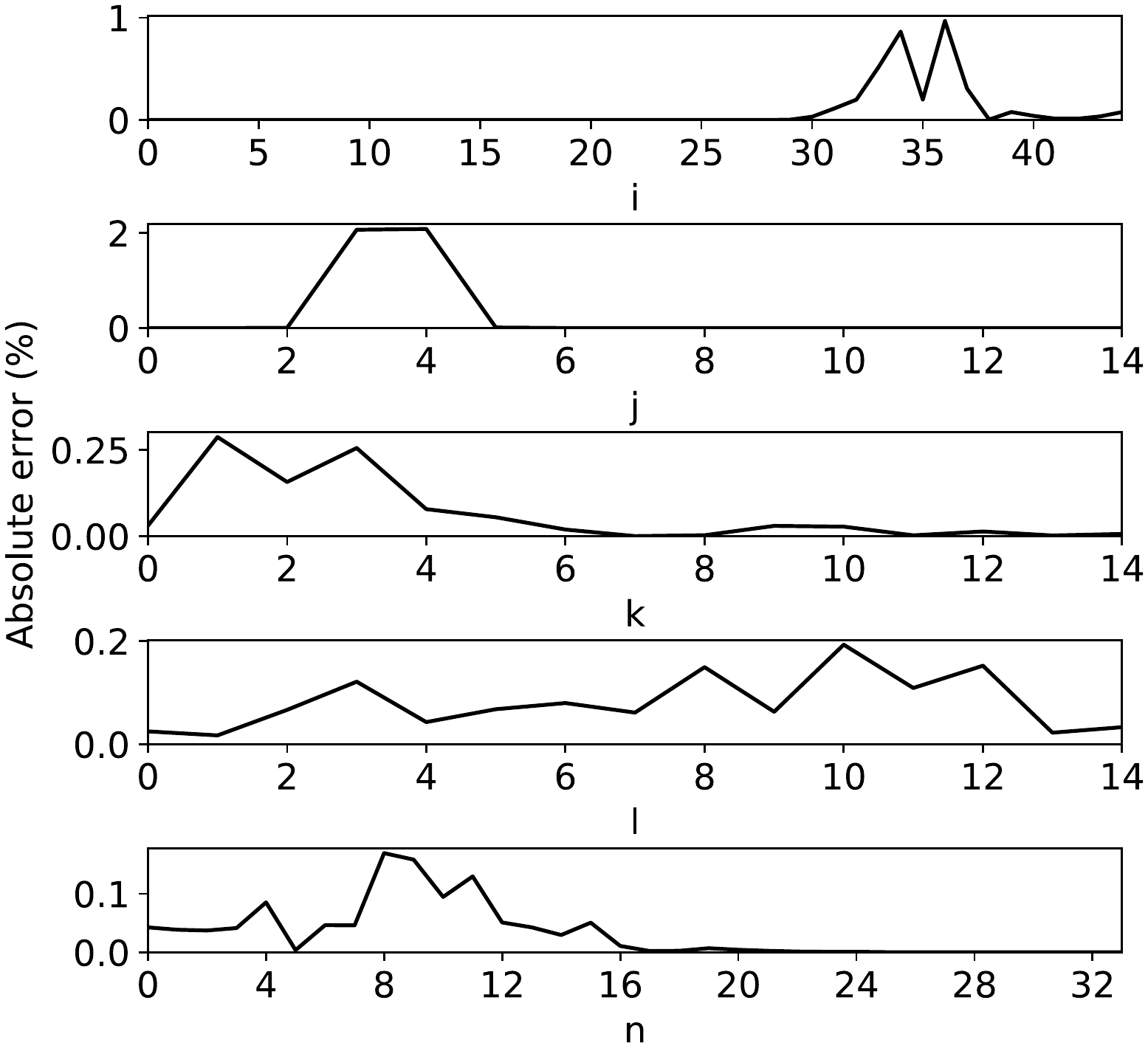}
  \caption{
    Absolute error between the probability distributions presented in \figref{fig:appli:DPSvsRW:cmp}.\label{fig:appli:DPSvsRW:diff}
  }
\end{figure}

While the older mass yields rely only on the mass asymmetry coordinate, $\alpha_{\rm g}$, it is not sufficient to compare only this variable at scission as there could be changes in the distribution of other coordinates. 
Figure \ref{fig:appli:DPSvsRW:cmp} shows the distribution of the scissioned configurations in the full collective coordinate lattice space, $(i,j,k,l,n)$, between the standard random walk method and the method presented in this work. 
Despite the statistical versus deterministic aspect of these two approaches, both algorithms are in exceedingly good agreement. 
The absolute error between the two approaches is shown in Fig.~\ref{fig:appli:DPSvsRW:diff}. 
We find that the statistical nature of the DRW algorithm leads to a maximum of $\sim2$\% error in the distribution of the scission neck radius, while the statistical error in $\alpha_{\rm g}$ is the lowest of all the coordinates, on the order of 0.1\%. 
These two figures show that we are successfully able to reproduce past work with our new technique and that older works indeed have quantitatively very good estimates of the mass yields within the context of FRLDM so long as a large number of fission events are calculated. 


\section{Conclusion}
\label{sec:conclusion}

With this work, we have improved the quality and predictive power of the mac-mic method in several areas. First, we have enhanced the quality of the nuclear PES by removing spurious continuum effects in our five-dimensional finite-range model. Further, our new resolution procedure of the Lipkin-Nogami equations enables the description of pairing effects with very high accuracy. We refer to these improvements colloquially as the `Enhanced Finite-Range Liquid-Drop Model' or eFRLDM for short. Second, our new deterministic algorithm, named DPS, completely removes statistical uncertainties when computing the fission fragment distribution of a particular nucleus. DPS enables the starting point of our calculation to be located at the ground state (easily identifiable for all nuclei) without the requirement of including a biased potential that artificially tilts the PES. We have defined a high-accuracy convergence criterion associated with our algorithm that affords ability to monitor the error associated with the obtained results. Last, but not least, we have generalized the particle number projection technique introduced for independent quasi-particle states in Ref.~\cite{simenel2010particle, scamps2015superfluid} and calculated scissioned configurations with it. This projection technique allows for the calculation of the coupled fragment charge and mass yield, $Y(Z,A)$. 

Our first eFRLDM+DPS results are presented for the pre-neutron fission fragments probability distributions of the reactions \textsuperscript{233,235}U(n,f) at different excitation energies. Our method can reproduce the odd-even staggering in the charge yields as well as the charge polarization of the fragments without any additional free parameters in the model. We find that a charge misalignment exists between the light and heavy fragments for these two reactions on the order of $\Delta_Z > 3.5$ charge units. Our results show a remarkable quantitative agreement with experimental data for charge yields and isobaric charge yields. We further highlight the capacity to reproduce past work within the context of our new methodology. 

Looking forward, our description of the temperature dependence of the PES is treated from a macroscopic perspective in this work. Since microscopic effects are relevant, especially at low incident energies, it might change the relative contribution of different saddles and impact our results. Also, in this work, we have focused primarily on pre-neutron fragments yields. However, due to the timescale of prompt particle emission, there is no such experimental data to compare directly. Hence, we can only compare with post-neutron yields, which induces further assumptions and models. A next step in this latter direction is therefore to pursue simulating the de-excitation of the nascent fragments. Improvements can be made to the description of microscopic temperature dependence, for example, by including finite-temperature effects directly into the microscopic states at each point of the PES, and this will assist in addressing the excitation energy dependence of our yield predictions. Yet another planned improvement is to refit the parameters of the model to account for the changes in the size of the basis and Strutinsky method. 

The eFRLDM+DPS approach developed in this work is an ideal tool for large-scale precision calculations of fission fragment distributions required to model a range of phenomena, especially in astrophysical scenarios where it is important to obtain both charge and mass yields simultaneously to determine the relative abundances of lighter species. In addition to upcoming model improvements, we plan to study the application of our yields in a series of future efforts. 

\section*{Acknowledgments}

The authors would like to thank Nicolas Schunck for his advice, his careful reading of the manuscript and his helpful comments, as well as David Regnier, Patrick Talou, J{\o}rgen Randrup, Toshihiko Kawano, Ionel Stetcu, Arnie Sierk, and Peter M{\"o}ller for valuable discussions over the years.

M.V. and M.R.M. were supported by the US Department of Energy through the Los Alamos National Laboratory. 
Los Alamos National Laboratory is operated by Triad National Security, LLC, for the National Nuclear Security Administration of U.S. Department of Energy (Contract No. 89233218CNA000001). 
M.V. and M.R.M. were partly supported by the Fission In R-process Elements (FIRE) topical collaboration in nuclear theory, funded by the U.S. Department of Energy and through the Los Alamos Laboratory Directed Research \& Development Exploratory Research project entitled ``a dynamical approach to low-energy fission''. This work was partly performed under the auspices of the U.S Department of Energy by Lawrence Livermore National Laboratory under Contract DE-AC52-07NA27344.

\appendix

\section{Finite-Range Liquid Drop Model (FRLDM)}
\label{app:FRLDM}

We group here the main formulas and parameters that define our model that we use to calculate the nuclear potential energy surface. 
The macroscopic liquid-drop energy for even-even nuclei is
\begin{equation}
\begin{aligned}
  E_{\rm mac}(\MQS{\chi}) &=
       M_{\rm H} Z + M_{\rm n} N \\
    &- a_{\rm v}(1 - \kappa_{\rm v}I^2)A \\
    &+ a_{\rm s}(1 - \kappa_{\rm s}I^2)B_1(\MQS{\chi})A^{2/3} \\
    &+ a_0A^0B_{\rm W}(\MQS{\chi}) \\
    &+ c_1\frac{Z^2}{A^{1/3}}B_3(\MQS{\chi})\\
    &- c_4\frac{Z^{4/3}}{A^{1/3}}\\
    &+ f(k_{\rm f}r_{\rm p})\frac{Z^2}{A}\\
    &- c_{\rm a}(N-Z)\\
    &+ W|I|B_{\rm W}(\MQS{\chi})\\
    &- a_{\rm el}Z^{2.39}.
\end{aligned}
\end{equation}
In this expression, $A$, $Z$, $N$ are respectively the number of nucleons, protons and neutrons and $I$
is the relative neutron excess,
\begin{equation}
  I = \frac{N-Z}{A}.
\end{equation}
We note that the pairing term is zero for even-even nuclei, and thus does not appear in the formula. 

We shift globally the energy the PES such that the energy of the spherical point is zero. In this case, only the shape-dependent terms contribute to the PES. There are four such terms in our approach:
\begin{enumerate}
    \item the surface energy, calculated assuming a finite-range Yukawa-plus-exponential nuclear interaction with no folding~\cite{krappe1979unified},
    \item the Coulomb term, defined with a Yukawa folding of the sharp macroscopic density~\cite{davies1977calculation},
    \item the $A_0$ energy from Ref.~\cite{moller2016nuclear},
    \item the Wigner term from Ref.~\cite{moller2016nuclear}.
\end{enumerate}

By setting
\begin{equation}
    \fvec{\sigma}_a = \frac{|\fvec{r}-\fvec{r}'|}{a}
\end{equation}
and letting $\shapeInside{}$ represent the shape of the liquid drop whose volume is held constant and equal to $\frac{4}{3}\pi A {r_0}^3$, the shape-dependent energies are the relative surface energy associated with a Yukawa-plus-exponential finite-range two-body interaction,
\begin{equation}
    B_1(\MQS{\chi}) =
      -\frac{A^{-2/3}}{8\pi^2{r_0}^2a^4}
      \iint_{\rm V}
        \left[\fvec{\sigma}_a-2\right]
        \frac{e^{-\fvec{\sigma}_a}}{\fvec{\sigma}_a}
        \diff{\fvec{r}}\diff{\fvec{r}'} \ ,
\end{equation}
the relative Coulomb energy of a folded-Yukawa macroscopic density,
\begin{multline}
    B_3(\MQS{\chi}) =
      \frac{15A^{-5/3}}{32\pi^2a_{\rm den}{r_0}^5}
      \iint_{\rm V}
        \frac{%
          \diff{\fvec{r}}\diff{\fvec{r}'}%
        }{%
          \fvec{\sigma}_{a_{\rm den}}%
        }
        \Bigg[
          1 - \\
            \left(
              1 + \frac{\fvec{\sigma}_{a_{\rm den}}}{2}
            \right)e^{-\fvec{\sigma}_{a_{\rm den}}}
        \Bigg] \ ,
\end{multline}
and the shape-dependency of the $A_0$ and Wigner terms is defined as
\begin{equation}\label{eq:pes:BWdef}
    B_{\rm W}(\MQS{\chi}) =
      \begin{cases}
        \left(1 - \frac{S_3}{S_1}\right)^2 a_{\rm d}+1
          & \text{if there is a neck} \\
        \qquad 1
          & \text{otherwise.}
      \end{cases}
\end{equation}
In the last expression, $S_1$ is the area of the maximum cross section of the smaller one of the end bodies and $S_3$ is the area of the geometric shape $\MQS{\chi}$ at the neck location. The definition of $B_{\rm W}$ is slightly different than in Refs.~\cite{moller2016nuclear, mumpower2020primary} where the condition is only relative to the MQS parameter, $\sigma_2$. However, $\sigma_2<0$ does not imply the presence of a neck, defined as the existence of a local minimum in the sharp macroscopic density along the $z$-axis. Defining $l_1$, $l_2$ and $l_3$ as the respective centers of the left, middle and right bodies of the MQS shape, in the case where $l_2 \leq l_1$ or $l_3 \leq l_2$, the shape cannot exhibit a neck whatever the sign of $\sigma_2$. In this situation, $B_{\rm W}$ defined as in older work is not continuous at $\sigma_2=0$. Our new definition prevents this situation.

The model parameters we have used for the calculation of the PES are often referred to as \texttt{FRLDM2002} which correspond to the model parameters introduced in Ref.~\cite{moeller1995nuclear} with additional corrections of Ref.~\cite{moller2004five}. We present here only the parameters having an influence on the shape-dependent terms of the PES. The values of the fundamental constants we have used, taken from~\cite{moller2004five}, are presented in Table~\ref{tab:appli:param:FRLDM1992-fond}.
\begin{table}[b]
    \centering
    \begin{ruledtabular}
    \begin{tabular}{ccc}
      Parameter     & Value & Unit \\
      \hline
      $e_2        $ & 1.4399764   & MeV fm \\
      $m_{\rm amu}$ & 931.4943335 & MeV/c\textsuperscript{2} \\
      $m_{\rm nuc}$ & 1.007970689 & amu    \\
      $m_{\rm n}  $ & 1.008664891 & amu    \\
      $m_{\rm p}  $ & 1.007276487 & amu    \\
    \end{tabular}
    \end{ruledtabular}
    \caption{Truncation of the fundamental constants.}
    \label{tab:appli:param:FRLDM1992-fond}
\end{table}
Table~\ref{tab:appli:param:FRLDM1992-mac} references the parameters associated with the macroscopic part of the energy, taken from~\cite{moeller1995nuclear}.
\begin{table}[b]
    \centering
    \begin{ruledtabular}
    \begin{tabular}{ccc}
      Parameter     & Value & Unit \\
      \hline
      $r_0        $ & 1.16      & fm \\
      $a          $ & 0.68      & fm \\
      $a_{\rm den}$ & 0.70      & fm \\
      $W          $ & 30.0      & MeV
    \end{tabular}
    \end{ruledtabular}
    \caption{Part of the macroscopic parameters used in our approach to describe the fission process from~\cite{moeller1995nuclear}.}
    \label{tab:appli:param:FRLDM1992-mac}
\end{table}
Table~\ref{tab:appli:param:FRLDM2002-mac} reports the macroscopic parameters taken from~\cite{moller2004five}.
\begin{table}[b]
    \centering
    \begin{ruledtabular}
    \begin{tabular}{ccc}
      Parameter     & Value & Unit \\
      \hline
      $a_{\rm d}$   & 0.9       &     \\
      $a_{\rm s}$   & 21.33000  & MeV \\
      $\kappa_{\rm s}$ & 2.378  & MeV \\
      $a_0      $   & 2.04000   & MeV
    \end{tabular}
    \end{ruledtabular}
    \caption{Part of the macroscopic parameters used in our approach to describe the fission process from~\cite{moller2004five}.}
    \label{tab:appli:param:FRLDM2002-mac}
\end{table}

The parameters associated with the potential $V(\fvec{r};\MQS{\chi})$ defined in~\eqref{eq:appr:pes:micV} are listed in Table~\ref{tab:appli:param:FRLDM2002-mic}.
\begin{table}[t]
    \centering
    \begin{ruledtabular}
    \begin{tabular}{ccc}
      Parameter     & Value & Unit \\
      \hline
      $V_{\rm s}  $ & 52.5      & MeV \\
      $V_{\rm a}  $ & 48.7      & MeV \\
      $A_{\rm den}$ & 0.82      & fm  \\
      $B_{\rm den}$ & 0.56      & fm\textsuperscript{2}\\
      $a_{\rm pot}$ & 0.8       & fm  \\
      $k_{\rm p}  $ & 0.025     &     \\
      $l_{\rm p}  $ & 28.0      &     \\
      $k_{\rm n}  $ & 0.01875   &     \\
      $l_{\rm n}  $ & 31.5      &     \\
      $a_1        $ & 15.677    & MeV \\
      $a_2        $ & 22.00     & MeV \\
      $J          $ & 35.0      & MeV \\
      $L          $ & 99.0      & MeV \\
      $Q          $ & 25.0      & MeV \\
      $K          $ & 300.0     & MeV
    \end{tabular}
    \end{ruledtabular}
    \caption{Microscopic parameters associated with the potential~\eqref{eq:appr:pes:micV}.}
    \label{tab:appli:param:FRLDM2002-mic}
\end{table}

\section{Resolution of the Lipkin-Nogami equations}
\label{app:LNsolve}

When using a seniority-pairing interaction, the Lipkin-Nogami equations associated with a valence space of $N_{\rm v}$ energy levels are the set of $2N_{\rm v}+3$ nonlinear equations~\eqref{eq:pes:LN0}-\eqref{eq:pes:LNlambda2} with the same number of unknowns $v_k$, $\epsilon_k$, $\Delta$, $\lambda$ and $\lambda_2$. Some of these equations are associated with high-derivatives. To reduce the amplitude and the number of non-zeros derivatives, we substitute $u_k$, $v_k$ and $\epsilon_k$ by the variable $x_k$ and $\theta_k$ according to 
\begin{align}
  u_k &= \cos(\theta_i) \\
  v_k &= \sin(\theta_i) \\
  x_k &= \epsilon_k - \lambda.
\end{align}
The Lipkin-Nogami equations can then be rewritten and reorganized as
\begin{equation}
    \fvec{F}(\fvec{p}) = 0,
\end{equation}
where, setting $\bar{k} = k - L_{\rm min}$,
\begin{align}
  F_{2N_{\rm v}}(\fvec{p}) &=
    L_{\rm min}
  - N_{\rm pair}
  + \sum_{k=L_{\rm min}}^{L_{\rm max}}
      \sin^2(\theta_k)
  \label{eq:LNapx:F1}\\
  F_{2N_{\rm v}+2}(\fvec{p}) &=
    \left[
      \sum_{k=L_{\rm min}}^{L_{\rm max}}
        \frac{1}{\sqrt{{x_k}^2+\Delta^2}}
    \right]
  - \frac{2}{G}
  \label{eq:LNapx:F2}\\
  F_{2\bar{k}+1}(\fvec{p}) &=
    \frac{1}{2}
    \left[
      1
    - \frac{x_k}{\sqrt{{x_k}^2+\Delta^2}}
    \right] - \sin(\theta_k)^2
  \label{eq:LNapx:F3}\\
  F_{2\bar{k}}(\fvec{p}) &=
    (4\lambda_2-G)\sin(\theta_k)^2
  + e_k
  - x_k - \lambda
  \label{eq:LNapx:F4}\\
  F_{2N_{\rm v}+1}(\fvec{p}) &=
    A(\fvec{\theta})\lambda_2
  - \frac{G}{4}B(\fvec{\theta})
  \label{eq:LNapx:F5},
\end{align}
where the notations
\begin{multline}
  A(\fvec{\theta}) =
    \left[
      \sum_{k=L_{\rm min}}^{L_{\rm max}}
        \cos(\theta_k)^2\sin(\theta_k)^2
    \right]^2 \\
  - \sum_{k=L_{\rm min}}^{L_{\rm max}}
      \cos(\theta_k)^4\sin(\theta_k)^4
 \end{multline}
 \begin{multline}
  B(\fvec{\theta}) =
    \left[
      \sum_{k=L_{\rm min}}^{L_{\rm max}}
        \cos(\theta_k)^3\sin(\theta_k)
    \right] \times \\
    \left[
      \sum_{k=L_{\rm min}}^{L_{\rm max}}
        \cos(\theta_k)\sin(\theta_k)^3
    \right] \\
  - \sum_{k=L_{\rm min}}^{L_{\rm max}}
      \cos(\theta_k)^4\sin(\theta_k)^4,
\end{multline}
and
\begin{equation}
    \fvec{p} =
      \Big(
        \theta_{L_{\rm min}},x_{L_{\rm min}},
        \dots,
        \theta_{L_{\rm max}},x_{L_{\rm max}},
        \lambda,\lambda_2,\Delta\Big).
\end{equation}

The analysis of the dependencies of each equations leads to a maximum of $10N_{\rm v}+2$ non-vanishing elements in the Jacobian matrix $J_{\fvec{F}}(\fvec{p})$. Also, $J_{\fvec{F}}{\fvec{p}}$ is \textit{block-arrowhead}, which means that
\begin{equation}
    J_{\fvec{F}} =
      \begin{pmatrix}
        A & B \\
        C & D
      \end{pmatrix},
\end{equation}
where $A$ is a block-diagonal matrix. In our case, the blocks of $A$ are 2-dimensional matrices
\begin{equation}
    A_{\bar{k}} =
      \begin{pmatrix}
        \frac{
          \partial F_{2\bar{k}}
        }{
          \partial \theta_k
        }
          & \frac{
              \partial F_{2\bar{k}}
            }{
              \partial x_k
            } \\
        \frac{
          \partial F_{2\bar{k}+1}
        }{
          \partial \theta_k
        }
          & \frac{
              \partial F_{2\bar{k}+1}
            }{
              \partial x_k
            }
      \end{pmatrix}.
\end{equation}
The block-column matrix $B$ and row-column matrix $C$ are respectively associated with the following $2\!\times\! 3$-dimensional and $3\!\times\! 2$-dimensional blocks
\begin{align}
    B_{\bar{k}} &=
      \begin{pmatrix}
        \frac{
          \partial F_{2\bar{k}}
        }{
          \partial \lambda
        }
        & \frac{
            \partial F_{2\bar{k}}
          }{
            \partial \lambda_2
          }
          & 0 \\
        0
        & 0
          & \frac{
              \partial F_{2\bar{k}+1}
            }{
              \partial \Delta
            }
      \end{pmatrix} \\
    C_{\bar{k}} &=
      \begin{pmatrix}
        \frac{
          \partial F_{2N_{\rm v}}
        }{
          \partial \theta_k
        } & 0 \\
        \frac{
          \partial F_{2N_{\rm v}+1}
        }{
          \partial \theta_k
        } & 0 \\
        0
        & \frac{
            \partial F_{2N_{\rm v}+2}
          }{
            \partial x_k
          }
      \end{pmatrix},
\end{align}

In the following, we propose a method to solve the Lipkin-Nogami equations and more generally any system of equations associated with a \textit{block-arrowhead} Jacobian matrix at each point $\fvec{p}$ based on generalizations of the iterative Newton method. In our case, we use the cubic and the fifth-order iterative methods developed respectively by Homeier~\cite{homeier2004modified} and by Sharma \& Gupta~\cite{sharma2014efficient}. The idea of these methods is to improve the convergence properties of the Newton scheme by evaluating the Jacobian at different $\fvec{p}$. For example, in the Sharpa \& Gupta scheme, one step from iteration $i$ to $i+1$ is
\begin{align}
    \fvec{x}^{(i)} &\shortleftarrow
        \fvec{p}^{(i)}
      - \frac{1}{2}
          J_{\fvec{F}}(\fvec{p}^{(i)})^{-1}
            \fvec{F}(\fvec{p}^{(i)}) \\
    \fvec{y}^{(i)} &\shortleftarrow
        \fvec{p}^{(i)}
      - J_{\fvec{F}}(\fvec{x}^{(i)})^{-1}
          \fvec{F}(\fvec{p}^{(i)}) \\
    \fvec{p}^{(k+1)} &\shortleftarrow
        \fvec{y}^{(i)}
      - \Big[
          aJ_{\fvec{F}}(\fvec{x}^{(i)})^{-1}\nonumber\\
        &\qquad\qquad+ bJ_{\fvec{F}}(\fvec{y}^{(i)})^{-1}
        \Big]
        \fvec{F}(\fvec{y}^{(i)}),
\end{align}
where the fifth-order convergence is obtained when $a=2$ and $b=-1$. The Newton scheme is recovered by doing only the first step and $\fvec{p}^{(i+1)} \shortleftarrow \fvec{x}^{(i)}$, while the two first steps are present in the Homeier scheme and $\fvec{p}^{(i+1)} \shortleftarrow \fvec{y}^{(i)}$.

The blockwise inversion theorem gives the inverse of the Jacobian matrix as
\begin{equation}\label{eq:app:LNsolve:Jinv}
    \fvec{J}_{\fvec{F}}(\fvec{p})^{-1} = 
      \begin{pmatrix}
        A^{-1}+IM & -IT^{-1} \\
        -M        & T^{-1}
      \end{pmatrix},
\end{equation}
where
\begin{align}
    I &= A^{-1}B \\
    T &= D - CI \\
    M &= T^{-1}CA^{-1}.
\end{align}
Note that it is assumed that $A$ and $T$ are invertible. When it is not the case, we slightly perturb the diagonal elements of the non-invertible matrix.
This method requires inversion of $N_{\rm v}$ 2-dimensional matrices $A_{\bar{k}}$ and one 3-dimensional matrix $T$. However, the procedure gives a dense matrix. Instead, we directly calculate the four vectors
\begin{align}
    \label{eq:app:LNsolve:aJF}
    \fvec{a}^{(i)} &=
      J_{\fvec{F}}(\fvec{p}^{(i)})^{-1}
      \fvec{F}(\fvec{p}^{(i)}) \\
    \label{eq:app:LNsolve:bJF}
    \fvec{b}^{(i)} &=
      J_{\fvec{F}}(\fvec{x}^{(i)})^{-1}
      \fvec{F}(\fvec{p}^{(i)}) \\
    \label{eq:app:LNsolve:cJF}
    \fvec{c}^{(i)} &=
      J_{\fvec{F}}(\fvec{x}^{(i)})^{-1}
      \fvec{F}(\fvec{y}^{(i)}) \\
    \label{eq:app:LNsolve:dJF}
    \fvec{d}^{(i)} &=
      J_{\fvec{F}}(\fvec{y}^{(i)})^{-1}
      \fvec{F}(\fvec{y}^{(i)}),
\end{align}
such that
\begin{align}
    \fvec{x}^{(i)} &\shortleftarrow
        \fvec{p}^{(i)}
      - \frac{1}{2}
          \fvec{a}^{(i)} \\
    \fvec{y}^{(i)} &\shortleftarrow
        \fvec{p}^{(i)}
      - \fvec{b}^{(i)} \\
    \fvec{p}^{(k+1)} &\shortleftarrow
        \fvec{y}^{(i)}
      - \left[
          a\fvec{c}^{(i)}+ b\fvec{d}^{(i)}
        \right].
\end{align}
In the following, we note $\fvec{e}=\fvec{a}^{(i)},\fvec{b}^{(i)},\fvec{c}^{(i)}\text{ or }\fvec{d}^{(i)}$ and $\fvec{f}=\fvec{F}(\fvec{p}^{(i)})\text{ or }\fvec{F}(\fvec{y}^{(i)})$ according to the equation~\eqref{eq:app:LNsolve:aJF}-\eqref{eq:app:LNsolve:dJF} considered. By injecting~\eqref{eq:app:LNsolve:Jinv} in~\eqref{eq:app:LNsolve:aJF}-\eqref{eq:app:LNsolve:dJF} we obtain
\begin{align}
    \fvec{e}_0 &=
        A^{-1}\fvec{f}_0
      + IM\fvec{f}_0
      - IT^{-1}\fvec{f}_1 \\
    \fvec{e}_1 &=
      - M\fvec{f}_0
      + T^{-1}\fvec{f}_1,\label{eq:app:LNsolve:e1}
\end{align}
where
\begin{equation}
    \fvec{e} =
      \begin{pmatrix}
        \fvec{e}_0 \\ \fvec{e}_1
      \end{pmatrix} \qquad
    \fvec{f} =
      \begin{pmatrix}
        \fvec{f}_0 \\ \fvec{f}_1
      \end{pmatrix},
\end{equation}
$\fvec{e}_0$ and $\fvec{f}_0$ are vectors of dimension $2N_{\rm v}$ and $\fvec{e}_1$ and $\fvec{f}_1$ are vectors of dimension $3$. The expression of $\fvec{e}_1$ appears in the expression of $\fvec{e}_0$. Thus, once $\fvec{e}_1$ is obtained by using~\eqref{eq:app:LNsolve:e1}, $\fvec{e}_0$ can be obtained through the expression
\begin{equation}
    \fvec{e}_0 =
        A^{-1}\fvec{f}_0
      - I\fvec{e}_1.
\end{equation}

Lastly, our convergence criteria is set to $\varepsilon^{\rm LN}(\fvec{p})<10^{-10}$, where, using the aforementioned functions $F_k$,
\begin{equation}
    \varepsilon^{\rm LN}(\fvec{p}) = \max_k{\abs{F_k(\fvec{p})}} \ .
\end{equation}

\section{Non-zero probability for odd-number of particles}
\label{app:nonzeroodd}

At first sight, the possibility to obtain a non-zero probability associated with odd-number of particles in the fragments distribution could seem to be a paradox in the case of static mean-field calculations in the case of even-even systems since:
\begin{enumerate}
    \item the state describing the fissioning system is time-even;
    \item the operators $\hat{P}_X$ and $\hat{P}^{(L)}_{\Xfrag}$ are time-even;
    \item a state describing odd-number fragments cannot be time-even.
\end{enumerate}
However, this paradox is only apparent and can be solved by noticing that the projection operators are both acting on the full $X$-body wavefunction (for each isospin). Even in the case where $\Xfrag$ is odd, the state $\hat{P}^{(L)}_{\Xfrag}\hat{P}_X\ket{\Phi(\MQS{\chi})}$ contains an even number of particles: $\Xfrag$ in the left fragment and $X-\Xfrag$ in the right one. It is thus time-even. It can easily be seen from a simple example of a time-even state having only two particles
\begin{equation}
  \label{eq:appr:frag:def2parttot}
  \ket{\Phi} = \cPw{i}\cPw{\timerev{i}}\ket{0},
\end{equation}
where $\cPw{i}$ is the creation operator of a particle in state $i$, $\timerev{i}$ is the time-reveral state of $i$. Both states can be decomposed in a similar way as in~\cite{simenel2010particle,verriere2019number} as
\begin{align}
    \label{eq:appr:frag:defi}
    \cPw{i}           &=
        \alpha^{(L)}_{i} \cPw[L]{i}
      + \alpha^{(R)}_{i} \cPw[R]{i} \\
    \label{eq:appr:frag:deftri}
    \cPw{\timerev{i}} &=
        \alpha^{(L)\star}_{i} \cPw[L]{\timerev{i}}
      + \alpha^{(R)\star}_{i} \cPw[R]{\timerev{i}},
\end{align}
where $\cPw[L]{k}$ and $\cPw[R]{k}$ are respectively the left and right parts of $\cPw{k}$ for $k=i,\timerev{i}$. The creation operators on the left commute with the ones on the right due to the complete separation of their spatial domain, and each of them commute with their time-reversal. Therefore, by injecting~\eqref{eq:appr:frag:defi} and~\eqref{eq:appr:frag:deftri} into~\eqref{eq:appr:frag:def2parttot} and developing the resulting expression, we obtain
\begin{equation}
\begin{aligned}
    \ket{\Phi} &=
      \alpha^{(L)}_{i} \alpha^{(L)\star}_{i}
        \cPw[L]{i} \cPw[L]{\timerev{i}}
          \ket{0} \\
               &+
      \big(
        \alpha^{(L)}_{i} \alpha^{(R)\star}_{i}
          \cPw[L]{i} \cPw[R]{\timerev{i}} \\
                  &\qquad+
        \alpha^{(R)}_{i} \alpha^{(L)\star}_{i}
          \cPw[R]{i} \cPw[L]{\timerev{i}}
      \big)\ket{0} \\
               &+
      \alpha^{(R)}_{i} \alpha^{(R)\star}_{i}
        \cPw[R]{i} \cPw[R]{\timerev{i}}
          \ket{0}.
\end{aligned}
\end{equation}
The three terms are orthogonal to each other and are all time-even. The first one corresponds to a state with two particles in the left side and zero in the right one, the second one corresponds to one particle on each side and the last one corresponds to two particles in the right side and zero in the left one. Therefore, even though $\ket{\Phi}$ is time-even and $\hat{P}^{(L)}_1$ is time-even, we have a non-zero probability to have odd-number fragments when the state $i$ spreads on both the left and right domains
\begin{equation}
    \prob{X_{\rm L} = \Xfrag} =
    \frac{
      \braket[\hat{P}^{(L)}_{\Xfrag}]{\Phi}{\Phi}
    }{
      \braket{\Phi}{\Phi}
    } = 2|\alpha_i^{(L)}\alpha_i^{(R)}|^2.
\end{equation}
However, as illustrated in Ref.~\cite{verriere2019number} in the case of static time-even Bogoliubov states with an even-number of particles and as demonstrated in \subsecref{subsec:appli:results}, the probability associated with odd-number fragments collapses to zero as soon as the fragments are separated enough and don't interact anymore. This is a direct consequence of the finite-range character of the nuclear interaction and the minimization of the energy: if two subsystems $\mathcal{S}_1$ and $\mathcal{S}_2$ of a system $\mathcal{S}$ do not interact with each other, the energy of the total system is the sum of the energies of both subsystems and thus, the state that minimize the energy of $\mathcal{S}$ is the product of the states minimizing each subsystems.

\bibliography{biblio}

\end{document}